\documentclass[useAMS,usenatbib,breaklinks=true]{mnras}
\usepackage{graphicx}
\usepackage{times}
\usepackage{hyperref}
\usepackage{amssymb}
\usepackage{amsmath}
\usepackage{color}
\usepackage{eso-pic}

\begin{document}

\AddToShipoutPictureBG*{%
  \AtPageUpperLeft{%
    \hspace{0.75\paperwidth}%
    \raisebox{-2.5\baselineskip}{%
      \makebox[0pt][l]{\textnormal{DES 2017-0232}}
}}}%

\AddToShipoutPictureBG*{%
  \AtPageUpperLeft{%
    \hspace{0.75\paperwidth}%
    \raisebox{-3.5\baselineskip}{%
      \makebox[0pt][l]{\textnormal{FERMILAB-PUB-17-162-A-AE}}
}}}%

\title[Optimizing BAO Measurements for photoz surveys] 
{Optimized Clustering Estimators for BAO Measurements Accounting for Significant Redshift Uncertainty}

\author[A. J. Ross et al.]{\parbox{\textwidth}{
Ashley J. Ross\thanks{Email: ross.1333@osu.edu; Ashley.Jacob.Ross@gmail.com}$^{1,2}$, 
Nilanjan Banik$^{3,4}$, 
Santiago Avila$^{2,5,6}$,
Will J. Percival$^2$,
Scott Dodelson$^{4}$,
Juan Garcia-Bellido$^{5,6}$,
Martin Crocce$^7$,
Jack Elvin-Poole$^8$,
Tommaso Giannantonio$^{9,10}$,
Marc Manera$^{10,11}$,
Ignacio Sevilla-Noarbe$^{12}$
}
  \vspace*{4pt} \\ 
$^{1}$Center for Cosmology and AstroParticle Physics, The Ohio State University, Columbus, OH 43210, USA\\
$^{2}$Institute of Cosmology \& Gravitation, Dennis Sciama Building, University of Portsmouth, Portsmouth, PO1 3FX, UK\\
$^{3}$ Department of Physics, University of Florida, Gainesville, Florida 32611, USA\\
$^{4}$ Fermi National Accelerator Laboratory, Batavia, Illinois 60510, USA\\
$^{5}$Departamento de F\'isica Te\'orica, M\'odulo C-15, Facultad de Ciencias, Universidad Aut\'onoma de Madrid, 28049 Cantoblanco, Madrid, Spain\\
$^{6}$Instituto de F\'isica Te\'orica, UAM-CSIC, Universidad Autonoma de Madrid, 28049 Cantoblanco, Madrid, Spain\\
$^7$Institut de Ci\`encies de l'Espai, IEEC-CSIC, Campus UAB, Carrer de Can Magrans, s/n,  08193 Bellaterra, Barcelona, Spain\\
$^8$Jodrell Bank Center for Astrophysics, School of Physics and Astronomy, University of Manchester, Oxford Road, Manchester, M13 9PL, UK\\
$^9$Universit\"{a}ts-Sternwarte M\"{u}nchen, Fakult\"{a}t f\"{u}r Physik, Ludwig-Maximilians-Universit\"{a}t M\"{u}nchen, Scheinerstrasse 1, 81679 M\"{u}nchen, Germany\\
$^{10}$Kavli Institute for Cosmology, University of Cambridge, Institute of Astronomy, Madingley Road, Cambridge, CB3 0HA, UK\\
$^{11}$Centre for Theoretical Cosmology, Department of Applied Mathematics and Theoretical Physics, Wilberforce Road, Cambridge CB3 0WA, UK\\
$^{12}$Centro de Investigaciones Energ\'eticas, Medioambientales y Tecnol\'ogicas (CIEMAT), Madrid, Spain
}
\date{Accepted by MNRAS} 

\pagerange{\pageref{firstpage}--\pageref{lastpage}} \pubyear{2016}
\maketitle
\label{firstpage}

\begin{abstract}
We determine an optimized clustering statistic to be used for galaxy samples with significant redshift uncertainty, such as those that rely on photometric redshifts. To do so, we study the baryon acoustic oscillation (BAO) information content as a function of the orientation of galaxy clustering modes with respect to their angle to the line-of-sight (LOS). The clustering along the LOS, as observed in a redshift-space with significant redshift uncertainty, has contributions from clustering modes with a range of orientations with respect to the true LOS. For redshift uncertainty $\sigma_z \geq 0.02(1+z)$ we find that while the BAO information is confined to transverse clustering modes in the true space, it is spread nearly evenly in the observed space. Thus, measuring clustering in terms of the projected separation (regardless of the LOS) is an efficient and nearly lossless compression of the signal for $\sigma_z \geq 0.02(1+z)$. For reduced redshift uncertainty, a more careful consideration is required. We then use more than 1700 realizations (combining two separate sets) of galaxy simulations mimicking the Dark Energy Survey Year 1 sample to validate our analytic results and optimized analysis procedure. We find that using the correlation function binned in projected separation, we can achieve uncertainties that are within 10 per cent of  those predicted by Fisher matrix forecasts. We predict that DES Y1 should achieve a 5 per cent distance measurement using our optimized methods. We expect the results presented here to be important for any future BAO measurements made using photometric redshift data. 
\end{abstract}


\section{Introduction}

Measurement of the location of the baryon acoustic oscillation (BAO) feature in the clustering of galaxies has proven to be a robust and precise method to measure the expansion history of the Universe and the properties of dark energy (see, e.g.,\citealt{WeinbergDERev,Aubourg15,Ross16,VargasDR12BAO}). The most successful results have been obtained by galaxy redshift surveys such as the SDSS-III Baryon Oscillation Spectroscopic Survey \citep{Eis11,Dawson12,AlamBOSSDR12}. Such surveys obtain precise redshift information ($\sim$ 0.1 per cent) and therefore are able to resolve the BAO along and transverse to the line of sight, providing simultaneous measurement of the expansion rate, $H(z)$, and the angular diameter distance, $D_A(z)$.

Multi-band imaging surveys rely on photometric redshifts for radial information. The Dark Energy Survey (DES) and Large Synoptic Survey Telescope are two such surveys and will produce enormous imaging catalogs and will rely on photometric redshifts for their radial information and thus their BAO measurements. The precision that can be achieved for bright, red samples is typically 3 per cent (or better; \citealt{redmagic}); such precision is good enough to localize the galaxies for tomographic studies but removes most of the radial BAO information. The angular diameter distance information is depressed, but still measurable (c.f. \citealt{SeoEis03,BlakeBridle05,Zhan09} for forecasts and  \citealt{Pad07,Estrada09,Hutsi10,Sanchez11,Seo12,Carnero12} for measurements). 

The above references have focused on using angular clustering measurements in narrow redshift slices. As the available BAO information mainly affords a measurement of $D_A$, such a focus obtains most of the information. However, it clearly does not use the full information available, as the radial binning blends the data beyond that induced by the redshift error.  Furthermore, cross-correlations between redshift slices are often ignored for simplicity. Determining the angular and physical separations of all galaxy pairs and weighting the information should allow more precise BAO measurements. \cite{Estrada09} present an analysis to do so in the context of photometric galaxy clusters and \cite{Montero16} analyze the results one expects for surveys that obtain redshifts from narrow-band filters. \cite{Eth15,Eth17} have conducted similar studies on the effect of DES redshift uncertainties on measures of galaxy environment. Here, we primarily focus on the results for Gaussian\footnote{See, e.g., \cite{Asorey16} for a study on effects of non-Gaussian redshift uncertainties.} redshift uncertainties close to 0.03(1+z).

In this study, we develop a framework for an optimal BAO analysis of a large-scale structure sample in the presence of significant redshift uncertainty. The paper is outlined as follows: In Section \ref{sec:BAOinfo} we study the distribution of BAO information within a given volume and in the presence of redshift uncertainty. This section begins with a discussion of how redshift errors impact the observed signal, then investigates the signal to noise using Fisher matrix formalism, and finally derives optimal weighting as a function of the redshift uncertainty. In Section \ref{sec:mocks} we describe how we simulate 1700 realizations of the DES Year 1 BAO galaxy sample (mocks) and measure their clustering statistics. In Section \ref{sec:BAOmock} we describe how we measure the BAO scale in these mocks and the results of these measurements. We conclude in Section \ref{sec:disc} with a discussion summarizing the results and presenting a recommended approach. 

The fiducial cosmology we use for this work is flat $\Lambda$CDM with $\Omega_{\rm matter} = 0.25$. Such a low matter density is ruled out by current observational constraints (see, e.g., \citealt{Planck2015}). However, the cosmology we use is matched to that of the MICE \citep{CrocceMICE,FosalbaMICE,FosalbaMICE2,CarreteroMICE} $N$-Body simulation, which was used to calibrate the mock galaxy samples we employ to test and validate our methodology. Throughout, we will use a fiducial redshift of 0.8, as this is close to the effective redshift we expect for DES Y1 BAO studies.

\section{BAO Information}
\label{sec:BAOinfo}

In this section, we outline the factors that affect the distribution of BAO information. We begin by describing how the BAO signal changes as a function of the orientation of clustering modes with respect to their angle to the line-of-sight (LOS), and demonstrate the effect in configuration space in terms of the observed orientation. We then investigate how the signal to noise, given by the Fisher matrix prediction, varies as a function of the true and observed LOS orientation. Here, we consider the `true' LOS orientation to be that one would observe in the absence of redshift uncertainties (e.g., in an ideal galaxy redshift survey) and the observed LOS orientation being that observed due to redshift uncertainties (e.g., in a photometric redshift galaxy sample). In this formulation, the true (and the observed) LOS orientation is affected by redshift-space distortions. We test different assumptions about the number density, redshift uncertainty, and clustering amplitude within the sample. Finally, we discuss how to weight galaxies in any given volume, given that the number density, redshift uncertainty, and clustering amplitude are all likely to evolve significantly with redshift.

Defining a sphere around a given LOS, equal volume elements are defined by the cosine of the angle to the LOS, $\mu$. Thus, in real, non-evolving, space, information is divided evenly in bins of $\mu$. These facts make the separation, $s$, (or wave number $k$) and $\mu$ a convenient pair of variables to use when considering the information content of large-scale structure data. The component of any property along the LOS is denoted using $_{||}$ and that transverse is denoted using $_{\perp}$. In what follows, we will use $\mu$ to denote the true $\mu$ in redshift-space (with no redshift uncertainty) and $\mu_{\rm obs}$ the observed $\mu$, which differs from the true $\mu$ due to errors in the redshift determination.

\subsection{Configuration-space signal with respect to the line-of-sight}

\begin{figure*}
\begin{minipage}{7in}
\includegraphics[width=7in]{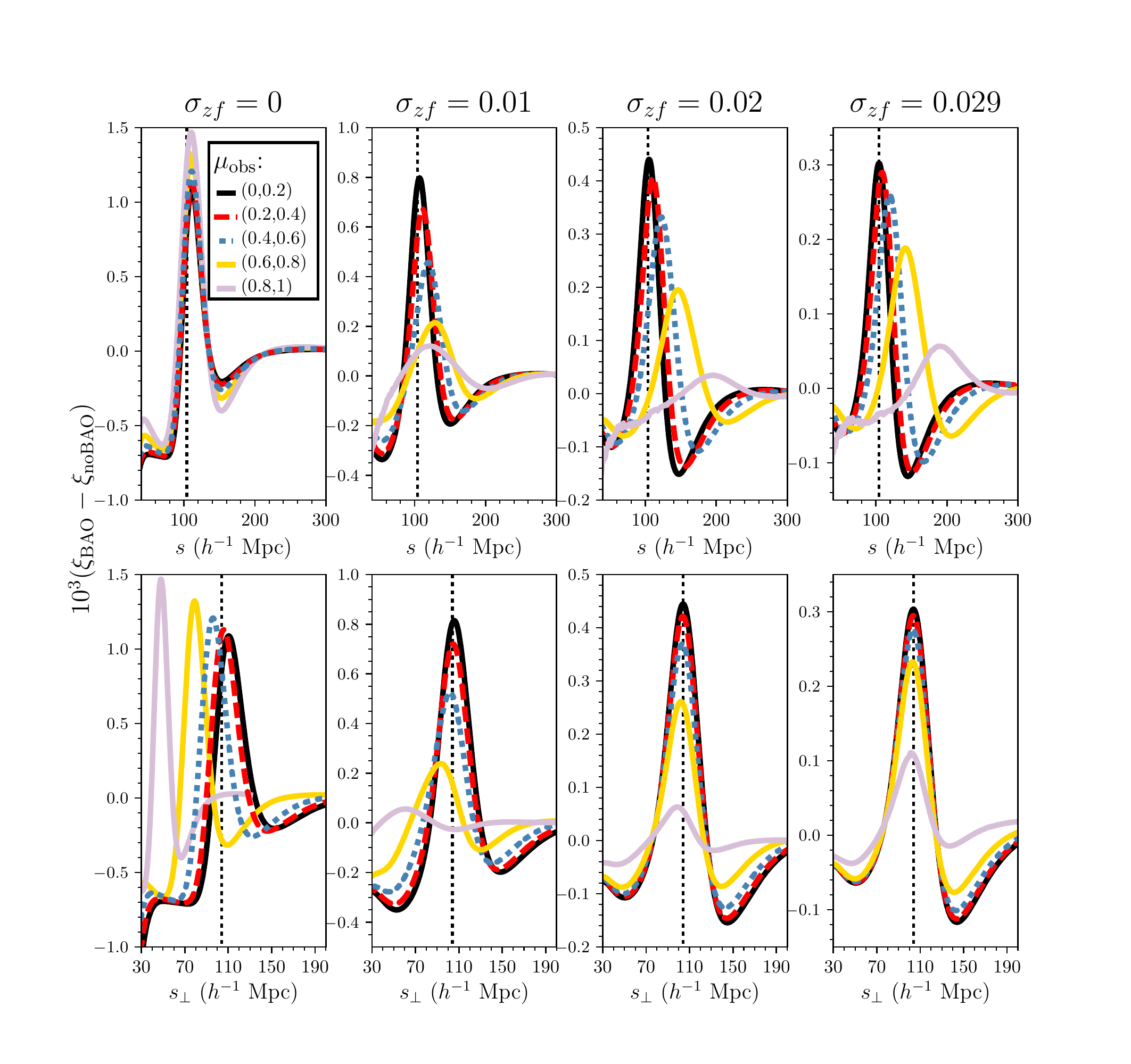}
  \caption{The top panels display the predicted BAO signal, separated by orientation of galaxy pairs with respect to the line of sight. This is orientation is parameterized by $\mu_{\rm obs}$, the cosine of the observed angle a pair of galaxies makes with respect to the line of sight (which differs from the true $\mu$ due to redshift errors). We assume a sample of galaxies at $z=0.8$, and a bias of 1.8, with $\sigma_{zf} = \sigma_z/(1+z)$ increasing from 0 to 0.029 from left to right. In order to isolate the BAO signal, a model constructed from a smooth power spectrum has been subtracted. The bottom panels display the same information, but against the transverse separation, $s_{\perp} = s\sqrt{1-\mu_{\rm obs}^2}$. }
  \label{fig:BAOsig6pan}
  \end{minipage}
\end{figure*}

Our BAO modeling approach is derived from that of \cite{Xu12,alph,Ross16}. We will model the BAO signal that one can observe in configuration space rather than in Fourier space, though the linear matter power spectrum, $P_{\rm lin}(k)$, is our required input. We first obtain $P_{\rm lin}(k)$ from {\sc Camb}\footnote{camb.info} \citep{camb} and fit for the smooth `no-wiggle' $P_{\rm nw}(k)$ via the \cite{EH98} fitting formulae. We account for redshift-space distortions (RSD) and non-linear effects via
\begin{equation}
P(k,\mu) = (1+\mu^2\beta)^2\left((P_{\rm lin}-P_{\rm nw})e^{-k^2\Sigma_{\rm nl}^2}+P_{\rm nw}\right),
\label{eq:pkmu}
\end{equation}
where $\mu = {\rm cos}(\theta_{\rm LOS})$ = $k_{||}/k$,
\begin{equation}
\Sigma^2_{\rm nl} = (1-\mu^2)\Sigma^2_{\perp}/2+\mu^2\Sigma^2_{||}/2,
\label{eq:damp}
\end{equation}

\noindent and $\beta\equiv f/b$. This factor is set based on the galaxy bias, $b$, and effective redshift of the sample we are modeling, with $f$ defined as the logarithmic derivative of the growth factor with respect to the scale factor. The factor $(1+\beta\mu^2)^2$ is the `Kaiser boost' \citep{Kaiser87}, which accounts for linear-theory RSD. The BAO damping parameters are fixed at $\Sigma_{||}= 10 h^{-1}$Mpc and $\Sigma_{\perp}= 6 h^{-1}$Mpc, as were used in \cite{Ross16} and are close to the expected values at $z=0.6$ given by \cite{SeoEis07,Seo16}. We study clustering at redshift $\sim$0.8, but our conclusions are not sensitive to the precise $\Sigma$ values and these values produce a good match between theory and simulation, as shown in Section \ref{sec:mockclus}.

Given $P(k,\mu)$, we determine the multipole moments
\begin{equation}
P_{\ell}(k) = \frac{2\ell+1}{2}\int_{-1}^1 d\mu P(k,\mu)L_{\ell}(\mu),
\end{equation}
where $L_{\ell}(\mu)$ are Legendre polynomials. These are transformed to moments of the correlation function, $\xi_{\ell}$, via
\begin{equation}
\xi_{\ell}(s) = \frac{i^{\ell}}{2\pi^2}\int dk k^2P_{\ell}(k)j_{\ell}(ks).
\end{equation}
We then use 
\begin{equation}
\xi(s,\mu) = \sum_{\ell}\xi_{\ell}(s)L_{\ell}(\mu).
\label{eq:xismu}
\end{equation}
In order to create `no BAO' models, we set $P_{\rm lin} = P_{\rm nw}$ in Eq. \ref{eq:pkmu}. We can thus isolate the BAO signal by using $\xi - \xi_{\rm no BAO}$ (as in, e.g., figure 3 of \citealt{AlamBOSSDR12}).

The preceding paragraph described the standard approach to BAO modeling, generally applied to the analysis of spectroscopic galaxy samples. We now extend these results under the assumption of significant redshift uncertainty in order to derive photometric redshift space clustering $\xi_{\rm phot}(s,\mu_{\rm obs})$. Given $\xi(s,\mu)$, we then convolve the results with the redshift uncertainty to convert them to $\xi_{\rm phot}(s,\mu_{\rm obs})$. In practice, this implies properly averaging over the probability distributions for $s_{\rm true}$ and $\mu_{\rm true}$ given the $s$ and $\mu_{\rm obs}$ observed in photometric redshift space. We use slightly simplified approach and consider all pairs to be centered on some effective redshift, $z_{\rm eff}$ (for which we use 0.8). The uncertainty on the separation in redshift of a pair of galaxies is $\sqrt{2}\sigma_z$; this provides a Gaussian probability distribution, $G(z)$, (centered on $z_{\rm eff}$) for a redshift representing the change in redshift separation between the observed and true separations. This change in redshift separation is converted to a change in radial separation given the difference between comoving distance to $z_{\rm eff}$ and to $z$ sampled from $G(z)$. At every observed $s,\mu_{\rm obs}$, we thus take the weighted mean of\footnote{In order to be explicit, we use $\mu_{\rm true}$ here.} $\xi(s_{\rm true},\mu_{\rm true})$, evaluated using Eq. \ref{eq:xismu}, assuming the observed $s,\mu_{\rm obs}$ are centered on $z_{\rm eff}$. In equation form
\begin{equation}
\xi_{\rm phot}(s,\mu_{\rm obs}) = \int {\rm d}zG(z)\xi(s_{\rm true},\mu_{\rm true}),
\label{eq:xip}
\end{equation}
where $G(z)$ is the aforementioned Gaussian likelihood of width $\sqrt{2}\sigma_z$ centered on $z_{\rm eff}$. Using $\chi(z)$ as the comoving distance to redshift $z$, one can derive,
\begin{equation}
s_{\rm true}(z,s,\mu_{\rm obs}) = \left([\mu_{\rm obs} s+\chi(z_{\rm eff})-\chi(z)]^2+(1-\mu_{\rm true}^2)s^2\right)^{1/2},
\end{equation}
\begin{equation}
\mu_{\rm true} = \left(\mu_{\rm obs} s+\chi[z_{\rm eff}]-\chi[z]\right)/s_{\rm true}
\end{equation}
and thus evaluate Eq. \ref{eq:xip}. Evolution in the redshift uncertainty can be taken into account by evaluating Eq. \ref{eq:xip} for each redshift uncertainty and taking the appropriately weighted mean of the results. Non-Gaussian redshift uncertainties can potentially be modeled through modifications in $G(z)$, but we leave this for future study.

We use the above to take averages over any given observed $\mu_{\rm obs}$ window to create any particular template:
\begin{equation}
\xi_{\rm phot,W}(s) = \int_0^1d\mu_{\rm obs} W(\mu_{\rm obs})\xi_{\rm phot}(s,\mu_{\rm obs}).
\label{eq:xiphot}
\end{equation}
The $W(\mu_{\rm obs})$ should be normalized to integrate to 1, e.g., we will consider `wedges' \citep{Kazin12,Kazin13} where $W(\mu_{\rm obs}) = 1/(\mu_{\rm obs,~max}-\mu_{\rm obs,~min})$ within a given $\mu_{\rm obs}$ range and 0 outside of it. 

In Fig. \ref{fig:BAOsig6pan}, we display $\xi_{\rm phot,W}(s)-\xi_{\rm phot,W,no BAO}(s)$ for models with redshift uncertainty increasing in panels from left to right. The left-most panels display the results assuming no redshift uncertainty, as would be appropriate for a spectroscopic survey. Each panel shows the results for five $\mu_{\rm obs}$ bins of thickness $\Delta \mu_{\rm obs} = 0.2$. The top panels show the model as a function of redshift-space separation, $s$. One can observe that in all cases with redshift uncertinaty the BAO signal is diluted as $\mu_{\rm obs}$ increases. This dilution relative to the $\mu_{\rm obs}<0.2$ case becomes greater
as the redshift uncertainty decreases (e.g., compare the heights of the yellow curves representing the $0.6 < \mu_{\rm obs} < 0.8$ signal). This is due to the fact that in all cases the majority of the information is at low true $\mu$ (as explained in the following subsection), but when the redshift uncertainty is greater, more of the information is transferred to high observed $\mu_{\rm obs}$. For the same reason, the BAO peak also appears at greater observed $s$ at greater $\mu_{\rm obs}$. This shift is more pronounced when the redshift uncertainty is greater. In the case with no redshift uncertainty, the signal is greatest at high $\mu$ due to the `Kaiser boost' from RSD.

The bottom panels of Fig. \ref{fig:BAOsig6pan} show the results in terms of the transverse component of the redshift-space separation, $s_{\perp}$. One can see that for $\mu_{\rm obs} < 0.8$ and $\sigma_z/(1+z) \geq 0.02$, the BAO signal is at a nearly constant $s_{\perp}$. For $\mu_{\rm obs} > 0.8$, the signal strength is greatly diminished. This suggests that for $\sigma_z/(1+z) \geq 0.02$, measuring $\xi_{\rm phot}(s_{\perp})$ and weighting the signal appropriately with $\mu_{\rm obs}$ should obtain a nearly optimal clustering estimator to use for BAO distance measurements. 

For $\sigma_z/(1+z) = 0.01$, the peaks do not line up well when displayed as a function of the transverse separation. This is indicative of a significant amount of signal coming from radial clustering information. Clearly, at this level of redshift uncertainty, the optimal separation to use and the interpretation of results in terms of $D_A$ and $H$ measurements needs to be carefully considered.

\subsection{Signal to noise with respect to the line-of-sight}
\label{sec:BAOsn}

We now use Fisher matrix formalism in order to investigate the distribution of BAO signal to noise. In this section, we employ the \cite{SeoEis07} Fisher matrix techniques and work in Fourier space (though we will later return back to configuration space). Given the linear matter power spectrum $P_{\rm lin}(k)$ (implicitly at redshift, $z$, with no RSD, as in Eq. \ref{eq:pkmu}), the relative amount of BAO information, $F$, as a function of $\mu$ can be modeled as depending on galaxy sample's bias, $b$, number density, $n$, and redshift uncertainty, $\sigma_z$. Suppressing factors (e.g., survey volume) that do not affect the distribution of information, the relevant equation from \cite{SeoEis07} is\begin{equation}
F(\mu) \propto \int dk \frac{k^2{\rm exp}[-2(k\Sigma_s)^{1.4}]{\rm exp}[-k^2(1-\mu^2)\Sigma^2_{\perp}-k^2\mu^2\Sigma^2_{||}]}{\left(b^2P_{\rm lin}(k)/P_{0.2}+1/[nP_{0.2}R(k,\mu)]\right)^2},
\label{eq:Fmu}
\end{equation}
where $\Sigma_s$ is $1/k_{\rm Silk}$ given in \cite{EH98}; which accounts for `Silk damping' \citep{Silk} in the BAO signature produced just after recombination. We use $\Sigma_s=8.38h^{-1}$Mpc, the value \cite{SeoEis07} use for the WMAP3 cosmology, which is a close match to our fiducial cosmology.\footnote{Throughout, when using \citealt{SeoEis07} Fisher matrix forecasts, we will use the WMAP3 parameterization.} $P_{0.2}$ is the value of the power spectrum at $k = 0.2 ~h$Mpc$^{-1}$, which includes the galaxy bias, so $b^2P_{\rm lin}(k)/P_{0.2} =b^2P_{\rm lin}(k)/b^2P_{\rm lin}(0.2) = P_{\rm lin}(k)/P_{\rm lin}(0.2)$ independent of the galaxy bias. 

The modulation of the power spectrum amplitude due to RSD and redshift uncertainty enters in the $R(k,\mu)$ term:
\begin{equation}
R(k,\mu) \equiv (1+\beta\mu^2)^2{\rm exp}(-k^2\mu^2\Sigma_z^2).
\end{equation}
The factor $(1+\beta\mu^2)^2$ is the same `Kaiser boost' \citep{Kaiser87} as in Eq. \ref{eq:pkmu}. The redshift uncertainty, $\sigma_z$, is accounted for with the second factor, with $\Sigma_z = \frac{c}{H(z)}\frac{\sigma_z}{(1+z)}$, where $c$ is the speed of light, and $H(z)$ is the Hubble parameter at redshift $z$. The factor in the numerator of Eq. \ref{eq:Fmu} ${\rm exp}[-k^2(1-\mu^2)\Sigma^2_{\perp}-k^2\mu^2\Sigma^2_{||}]$ adds an additional LOS dependence. This factor accounts for damping of the BAO feature due to non-linear structure growth. Again following \cite{SeoEis07}, but rearranging the factors, we use $\Sigma_{\perp} = 8.3\sigma_8D(z) h^{-1}$Mpc and $\Sigma_{||} = (1+f)\Sigma_{\perp}$, where $D(z)$ is the linear growth factor, normalized to be 1 at $z=0$.

\begin{figure}
\includegraphics[width=94mm]{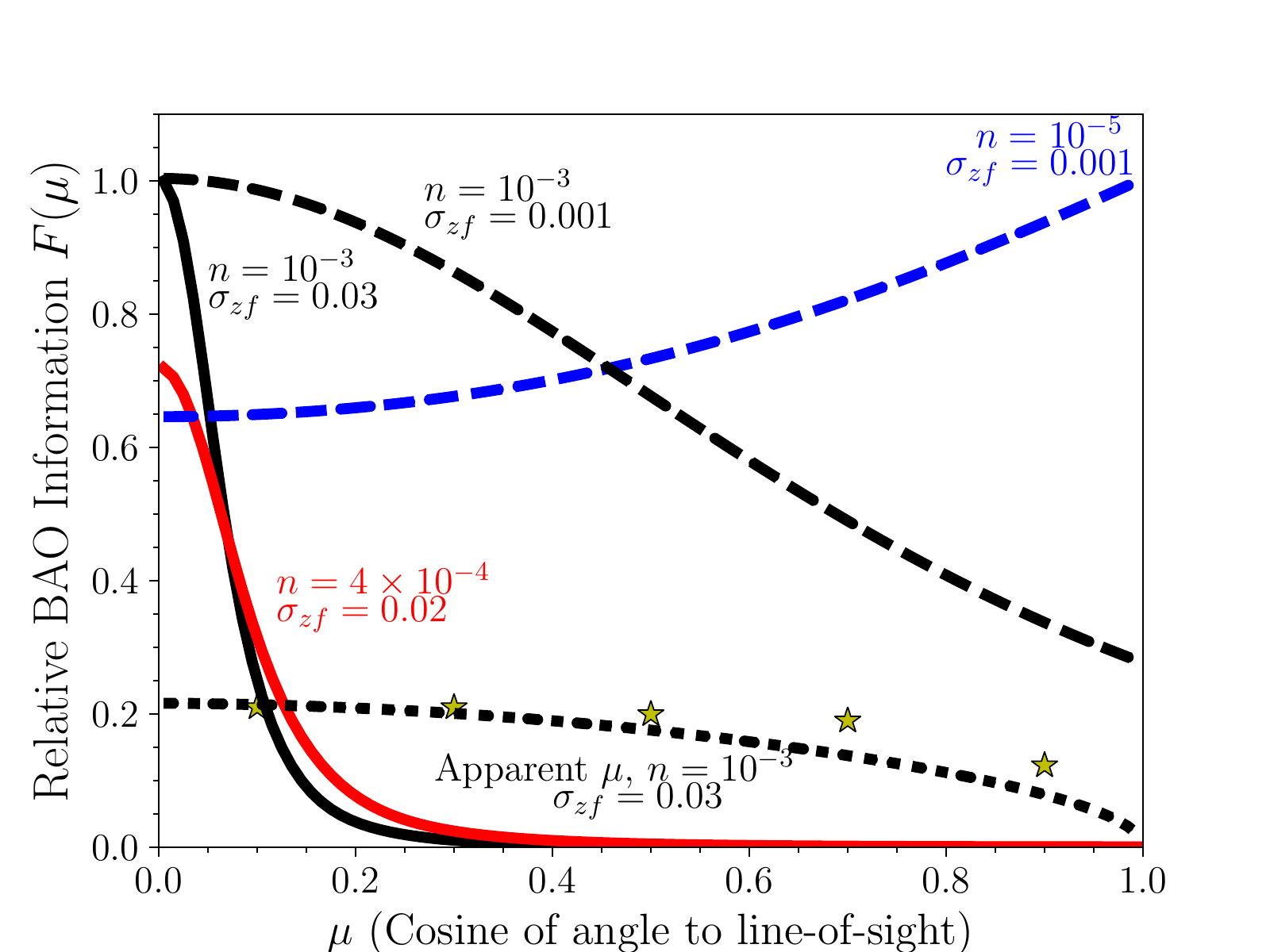}
  \caption{The relative amount of BAO information as a function of $\mu$, the cosine of the angle to the line of sight, for a sample of galaxies at $z=0.8$, with properties as labeled. We use $\sigma_{zf} = \sigma_z/(1+z)$ and the number density $n$ is in units $h^{3}$Mpc$^{-3}$. All cases use a galaxy bias of $b=1.8$, except for the solid red curve, which assumes $b=2$.
The dashed curves represent redshift precision achievable from spectroscopic galaxy surveys, while the solid curves show results achievable from multi-band imaging surveys, such as the Dark Energy Survey. All results are in terms of the true $\mu$, except for the dotted curve, which displays the information in terms of observed $\mu$. Results are normalized as follows: The solid black and red curves and the black dashed curve are all divided by the same factor; thus, they are directly comparable. The blue dashed curve is normalized such that it has the same maximum value as the black dashed curve; it thus illustrates how the relationship between BAO information and $\mu$ changes with number density. The black dotted curve displays the same information as the black solid line, but with respect to observed $\mu$; the area under the apparent $\mu$ curve is normalized to be twice that of the solid black curve (for legibility). The yellow stars represent what we find for mock galaxy samples, normalized so that the results match the dotted curve at $\mu=0.1$.}
  \label{fig:relinfo}
\end{figure}

There are thus three separate factors that are a function of $\mu$. Two are important even if there is no redshift uncertainty. The amplitude of the power spectrum is increased at high $\mu$ by RSD, thereby boosting the signal and thus the BAO information, while the BAO amplitude is damped to a greater extent at high $\mu$, thereby decreasing the signal and the amount of BAO information. These two factors are evident in Eqs. \ref{eq:pkmu} and \ref{eq:damp}. The degree to which one effect dominates over the other depends on the number density of the sample in question, which can be understood given the expectation for the variance of the power spectrum \citep{FKP}
\begin{equation}
\sigma^2_P \propto (P+1/n)^2.
\end{equation}
In the low number density limit, boosting the overall amplitude of the power spectrum is most important, while in the high number density limit, the fractional uncertainty on the measured power, $\sigma_P/P$, is constant and thus it is only the modulation of the BAO amplitude that matters. The competing effects are illustrated in Fig. \ref{fig:relinfo}. This figure displays the relative BAO information as a function of $\mu$ for multiple scenarios. The black and blue dashed curves show identical cases, except that the black curve is for a number density of $10^{-3}h^3$Mpc$^{-3}$ and the blue curve is for a number density of $10^{-5}h^3$Mpc$^{-3}$ (the dashed curves have been normalized so that the maximum value is the same for each, otherwise the $n=10^{-5}h^3$Mpc$^{-3}$ would have considerably smaller amplitude at all scales). 

The third factor that modulates the BAO information as a function of $\mu$ is the redshift uncertainty. In Fig. \ref{fig:relinfo}, we show the relative BAO information as a function of $\mu$ for two cases. The solid black curve displays the case where the number density is $10^{-3}h^3$Mpc$^{-3}$, the galaxy bias is 1.8, and the redshift uncertainty is $\sigma_z/(1+z) = 0.03$. These characteristics are similar to a sample of galaxies selected from DES Y1 data to be optimal for BAO measurements. The solid red curve displays the case where the number density is $4\times10^{-4}h^3$Mpc$^{-3}$, the galaxy bias is 2, and the redshift uncertainty is $\sigma_z/(1+z) = 0.02$. These characteristics are similar to the DES Y1 high luminosity `redmagic' \citep{redmagic} sample. We have normalized the black and red curves by the same factor (so that the black curve has a maximum value of 1) and thus the BAO information for these cases can be directly compared. The total BAO information (the area under each curve) matches to within 1 per cent. 

For each case representing DES Y1 data, the vast majority of the information is at low $\mu$; for $\sigma_z/(1+z) = 0.03$, half of the information is reached at $\mu=0.061$ while for $\sigma_z/(1+z) = 0.02$, half of the information is reached at $\mu=0.08$. The relative amount of information in $D_A$ compared to $H$ can be determined by integrating Eq. \ref{eq:Fmu} with respect to $\mu$ and weighting by a factor of $\mu^2$ for $H$ and $(1-\mu^2)$ for $D_A$ \citep{SeoEis07}. There is more than a factor of 100 times more information in $D_A$ than $H$ in the $\sigma_z/(1+z) = 0.03$ example, to be compared to a factor of 8/3 for the spherically symmetric case with no redshift uncertainty (see, e.g., \citealt{SeoEis07,Ross152D}). Put differently, using equations 9 and 10 of \cite{Ross152D} and the $\sigma_z/(1+z) = 0.03$ curve (solid black) in Fig. \ref{fig:relinfo} as their $F(\mu)$, the power-law indices are 0.99 for $D_A$ and 0.01 for $H$.

\begin{figure}
\includegraphics[width=84mm]{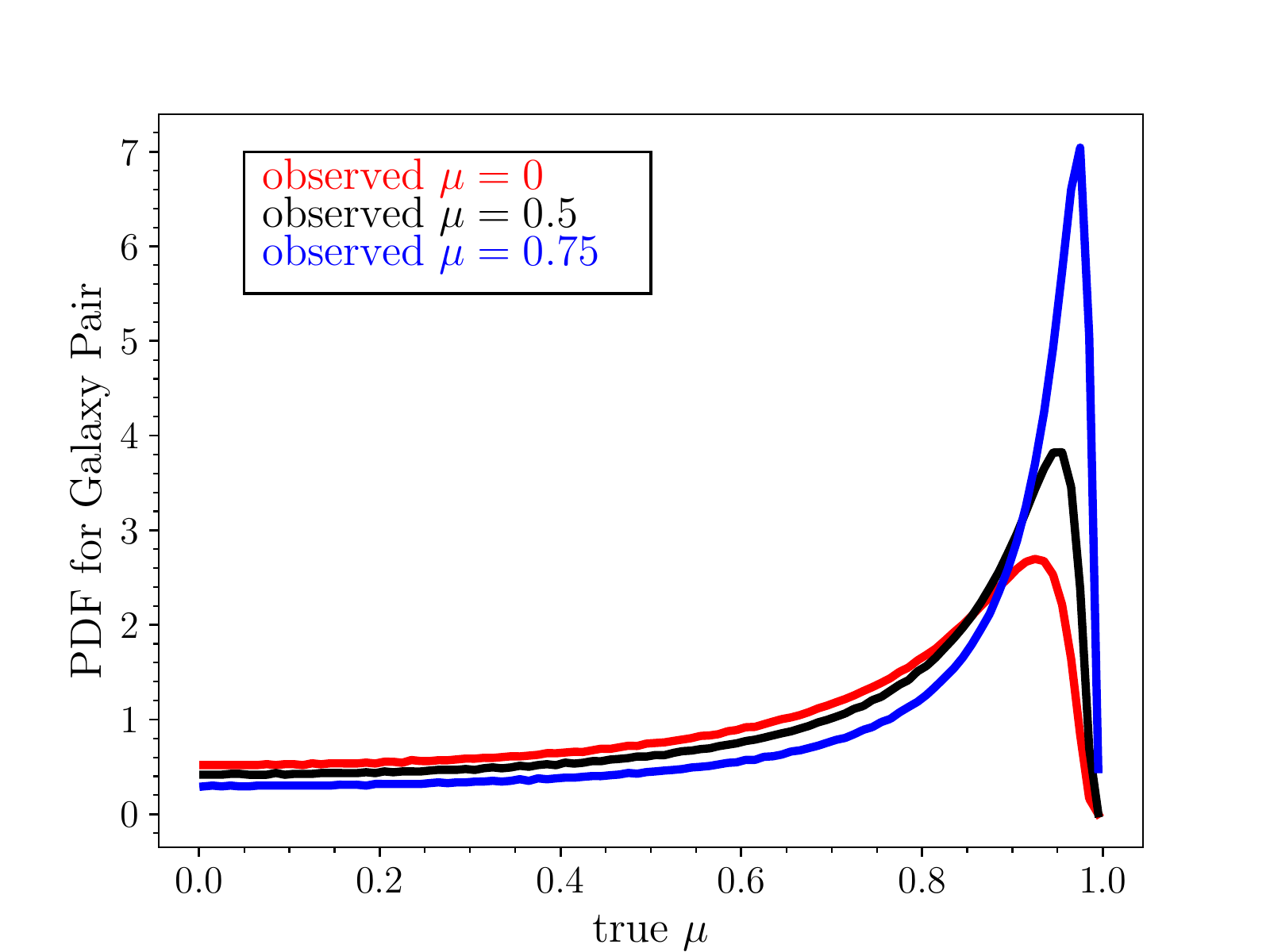}
  \caption{The configuration space mapping between observed and true $\mu$ for galaxy pairs of (true) separation 100$h^{-1}$Mpc and redshift uncertainty $\sigma_z = 0.03(1+z)$, centered at $z=0.8$. Each curve peaks at large $\mu$ due to the fact that the observed radial separation is dispersed by 157$h^{-1}$Mpc, on average.}
  \label{fig:muobstrue}
\end{figure}

Importantly, all curves on Fig. \ref{fig:relinfo} {\it except} for the dotted curve display the information in terms of the {\it true} $\mu$ distribution. This is implicit in the Fisher matrix formalism; power in the line of sight clustering modes is removed and this removes the information. However, this is the information with respect to the true orientation of the galaxies on the sky (i.e., assuming zero redshift uncertainty). The orientation we observe is in fact (strongly) affected by the redshift uncertainty, and we presented the effect of this on the measured signal in the previous subsection. In order to simulate how the signal to noise is distributed, we switch back to configuration space and determine the mapping between true and observed $\mu$ for galaxy pairs with separation $r=100h^{-1}$Mpc; we do this in the same manner as in Eq. \ref{eq:xip}. For each observed $\mu$, we have a probability distribution in true $\mu$, given the probability distribution for true redshift separation, itself given by the pair of observed redshifts and their uncertainties. We show this PDF for three observed $\mu$ cases in Fig. \ref{fig:muobstrue}, all for a sample with $\sigma_z = 0.054$, centered at $z=0.8$ (so at $z=0.8$, $\sigma_z = 0.03[1+z]$). One can see that for all observed $\mu$, the bulk of the pairs will be at high true $\mu$, but also a significant number will be from true $\mu\sim 0$, where the majority of the BAO information exists.

The dotted black curve in Fig. \ref{fig:relinfo} thus displays the $F(\mu)$ relative BAO information in the solid black curve, but in terms of {\it observed} $\mu$, accounting for the redshift uncertainty, using the mapping illustrated in Fig. \ref{fig:muobstrue}. The redshift uncertainty distributes the BAO information over a wide range in observed $\mu$. For example, for $\mu_{\rm obs} = 0$ the curve peaks at true $\mu=0.84$, due to the fact that at $z=0.8$ a redshift uncertainty of $0.03(1+z)$ implies a mean radial dispersion of 157$h^{-1}$Mpc (for our fiducial cosmology) and $157/\sqrt{157^2+100^2}=0.84$. The information is displayed so that the area under dotted curve is twice that of the solid curve (purely for legibility; they represent the same total amount of information). However, as the {\it true} information is represented by the solid curve, the information in the observed $\mu$ bins must be highly covariant (i.e., the BAO information in $\mu_{\rm obs} < 0.2$ will be highly covariant with that in $0.2 < \mu_{\rm obs} < 0.4$ since it must be coming from similar ranges in true $\mu$). This is a significant difference from the spectroscopic redshift-space scenario, where the information in different $\mu$ bins is close to independent \citep{Ross152D}. The correlation arises due to the fact that at a number density of 10$^{-3}h^3$Mpc$^{-3}$, the $\mu \sim 0$ clustering modes are significantly over-sampled, so that even when randomly distributing the information in observed $\mu$, one expects to find a significant correlation in the observed $\mu$ bins.

In \cite{Ross152D}, methodology was presented for obtaining optimal\footnote{\cite{Ross152D} used the variable $F$ for this window, but we use $W$ here so as not to confuse with the information $F$.} windows $W(\mu_{\rm obs})$ to be applied to clustering measurements, and modeled as in Eq. \ref{eq:xiphot}, in order to best extract the information on $D_A$ and $H$. Given the lack of information on $H$ for a realistic photometric redshift survey, discovering a single optimal weighting function in $\mu_{\rm obs}$ should provide an optimized BAO distance measure heavily weighted to $D_A$; the potential gain in defining multiple windows to obtain separated $D_A/H$ constraints is likely negligible. The dotted curve in Fig. \ref{fig:relinfo} represents an optimal $W(\mu_{\rm obs})$ window (ignoring the covariance at different $\mu_{\rm obs}$). However, there is an implicit assumption in \cite{Ross152D} that the BAO feature is nearly constant in $\mu$, in which it is trivial to combine information from any range in $\mu$ (e.g., given some optimal weighting function). In the preceding section, we investigated the signal as a function of $\mu_{\rm obs}$, showing the BAO signal to appear at the same $s_{\perp}$ independent of $\mu_{\rm obs}$. Thus, one expects the optimized clustering estimator for BAO from a photometric redshift survey to be $\int d\mu_{\rm obs} W_{\rm opt}(\mu_{\rm obs})\xi_{\rm phot}(s_{\perp},\mu)$, where $W_{\rm opt}(\mu_{\rm obs})$ can be obtained using the methods described in this section.

\subsection{Accounting for Variations in Redshift Uncertainty}

In the previous sub-section, we outlined how redshift uncertainty affects the distribution of BAO information with respect to the LOS orientation. In this sub-section, we study how one can take variations in redshift uncertainty within a given survey volume into account. Throughout, we assume Gaussian redshift uncertainties; studying the impact of non-Gaussian redshift uncertainties is a possible extension of these results.

\subsubsection{As a function of redshift}
\label{sec:zweight}
Generically, one expects that the number density, clustering amplitude, and redshift uncertainty will evolve with redshift for some selection of galaxies. In order to maximize signal to noise, one therefore wishes to properly weight the contribution from each volume. Given the variance on the power spectrum $\sigma_P^2 \propto (P+1/n)^2$, the square of the signal to noise ratio\footnote{We use $F_P$ to denote the square of the signal to noise ratio in a similar fashion as previous sections, but this now refers to power spectrum amplitude, rather than BAO, information.} , $F_P$, for each volume is $F_P \propto (nP)^2/(1+nP)^2$. Assuming the signal, $P$, is constant, this leads to the $k$ and $z$ dependent inverse-variance `FKP' weight per galaxy based on \cite{FKP}: 
\begin{equation}
w_{\rm FKP}(k,z) = \frac{1}{1+n(z)P_g(k)}.
\end{equation} 
$P_g(k)$ should be the measured amplitude of the power spectrum of the sample in question. Note that when changing from the inverse variance of the power to a per-galaxy weight, we divide by the number density $n$ in the numerator and we take the square root of $F_P$ due to the fact that the galaxy field is in effect squared when one obtains $P$.

For a sample with power that evolves as a function of redshift, we write $P_g = P_{\rm lin}b^2D^2$, where $P_{\rm lin}$ is the present day ($z=0$) linear theory power spectrum. We now have $F_P \propto (nP_{\rm lin}b^2D^2)^2/(1+nP_{\rm lin}b^2D^2)^2$. This leads to \citep{PVP}
\begin{equation}
w_{\rm FKP}(k,z) = \frac{b(z)D(z)}{1+n(z)P_{\rm lin}(k)b^2(z)D^2(z)},
\label{eq:pvp}
\end{equation} 
where one factor of $bD$ has been canceled due to the fact that each galaxy contributes its own $bD$ to the signal.

In practice, it is most useful to assign a single weight per galaxy, which requires evaluating $P_{\rm lin}(k)$ at some effective $k$, $k_{\rm eff}$. To find $k_{\rm eff}$, we take the weighted mean using the Fisher information at each $k$ value as the weight. The redshift uncertainty moves $k_{\rm eff}$ to lower values. For a number density $1.5\times10^{-3}h^{3}$Mpc$^{-3}$ and a bias of 1.8 at $z=0.65$, $k_{\rm eff}$ moves from 0.16$h$Mpc$^{-1}$ to 0.12$h$Mpc$^{-1}$ when comparing a sample with no redshift uncertainty to one with $\sigma_z =0.03(1+z)$.

In order to further account for redshift uncertainty, we recognize that in Eq. \ref{eq:Fmu}, at a given $k$ a change in the redshift uncertainty can be equivalently modeled as a change in the shot-noise term that is strongly dependent on the true $\mu$ (but more weakly dependent on the observed $\mu$, based on the results in the previous subsection). Thus, one can define an effective number density, $n_{\rm eff}$:
\begin{equation}
n_{\rm eff}(z) = n(z)\int d\mu R(k_{\rm eff},\mu). 
\label{eq:neff}
\end{equation}
This implies $F_P \propto (n_{\rm eff}P)^2/(1+n_{\rm eff}P)^2$. 
One can define a FKP weight per galaxy accounting for redshift uncertainty by simply substituting $n_{\rm eff}(z)$ for $n(z)$ and using $k_{\rm eff}$ to evaluate $P_{\rm lin}$
\begin{equation}
w_{\rm FKP}(z) = \frac{b(z)D(z)}{1+n_{\rm eff}(z)P_{\rm lin}(k_{\rm eff})b^2(z)D^2(z)}.
\label{eq:wfkpbz}
\end{equation} 
We have verified this weights galaxies approximately correctly by comparing to the forecasted uncertainty in redshift shells, from which a weight per galaxy can also be determined. 

As the redshift weights are simply inverse-variance weights, evaluating their expected impact is straight-forward. For instance, if half of the volume is occupied by galaxies have $w_{ \rm FKP}=1$ and the other half $w_{ \rm FKP}=0.5$, the variance {\it per galaxy} is twice as large for the sample with $w_{ \rm FKP}=0.5$. In this case, one can determine with a simple numerical experiment --- sampling Gaussians of the appropriate width --- that, assuming each volume has equal variance (so one half of the volume would have twice the total number of galaxies), the improvement in precision achieved from using the inverse-variance weights per galaxy is 5 per cent. If instead half the volume has a per galaxy weight of $w_{ \rm FKP}=0.1$, the improvement is 30 per cent.\footnote{ A notebook with these calculations is here: https://github.com/ashleyjross/LSSanalysis/tree/master/notebooks .}

\subsubsection{At a given redshift}
\label{sec:zpweight}
We have developed most of the tools and formalism required in order to probe the question of how to weight galaxies in a sample based on their photometric redshift uncertainty. The previous subsection derives weights assuming the redshift uncertainty can be treated as a constant at a given redshift. However, at a given redshift the photometric redshift uncertainty is likely to vary between galaxies based on their particular color or flux, we might therefore wish to weight galaxies differently based on these properties. This is analogous to how one might weight for varying galaxy bias within a population, as given by equation 28 of \cite{PVP}. 

Unfortunately, the dependence of Eq. \ref{eq:Fmu} on the redshift uncertainty is not as simple as the dependence on the galaxy bias. It is highly dependent on $\mu$; we must consider\footnote{ As in the previous subsection, we are ignoring the Kaiser boost factors.}
\begin{equation}
F_P  \propto \frac{\left(nP{\rm exp}(-k^2\mu^2\Sigma_z^2)\right)^2}{\left(1+nP{\rm exp}(-k^2\mu^2\Sigma_z^2)\right)^2}.
\end{equation}
Previously, we integrated over $\mu$ in order to obtain $n_{\rm eff}$. It is clear that now the relative weight as a function of redshift uncertainty would change considerably as a function of $\mu$. Thus, any optimal weighting is a function of $\Sigma_z$ {(which depends primarily on $\sigma_z$) and $\mu$ and it is not obvious how to apply such a weight on a per galaxy basis.\footnote{ A potential option is to consider a $\mu_{\rm eff}$ for each $\sigma_z$, but we do not investigate that.} 

Thus, to move forward we consider the possibility of two populations with different number density and redshift uncertainty.  How should the galaxies with the worse redshift uncertainty be weighted to give the optimal result? We are free to arbitrarily apply weights, $Z_i$, to each population when combining them. Doing so, we calculate the weighted number density, $n(z)_{\sigma_z}$, and a weighted-mean redshift uncertainty\footnote{Note, this is different than the unweighted mean quantity $\bar{\sigma}_z$ in Table \ref{tab:baosigz}.}, $\langle \sigma_{z}\rangle$, that we can enter into a Fisher matrix forecast (i.e., directly in to Eq. \ref{eq:Fmu}). For a general number of populations, the total number density will simply be $n_{\rm tot}=\sum n_i$; since we are considering a single redshift range, we drop the redshift dependence in what follows. If we are to weight each population, this must modulate the effective number density, $n_{\sigma_z}$, so that it is lower. 

{If we are to normalize the weight such that the mean weight is 1, i.e, $\sum^{i=N_{\rm tot}}_{i=0} n_i(\sigma_z)Z_i(\sigma_z) =n_{\rm tot}$, we can work out the effective number density, $n(z)_{\sigma_z}$, by considering Gaussian statistics. If one measures some parameter, X,  $N_{\rm tot}$ times using the exact same experiment, then $\langle X \rangle = 1/N_{\rm tot} \sum^{i=N_{\rm tot}}_{i=0} X_i$ and the uncertainty can be expressed by considering each element in the sum: $\sigma_{\rm tot}^{2} = 1/N^2_{\rm tot}\sum^{i=N_{\rm tot}}_{i=0} \sigma^2_{\rm d}$, given each measurement has the same variance $\sigma^2_{\rm d}$. Inspection reveals this leads to the standard
$\sigma_{\rm tot}^{-2} \propto N_{\rm tot}$. We wish to determine $N_{\rm eff}$ for the case where the result of each measurement is weighted differently. With $N_{\rm tot}$ measurements, our estimate is $\langle X_{\rm weighted} \rangle = 1/N_{\rm tot} \sum^{i=N_{\rm tot}}_{i=0} Z_i X_i$ (since we have chosen the weights to be normalized such that $\langle Z \rangle = 1$). 

The weighted uncertainty can be similarly written
$\sigma_{\rm tot,weighted}^{2} = 1/N^2_{\rm tot} \sum^{i=N_{\rm tot}}_{i=0} (Z_i\sigma_d)^{2} $. Defining $N_{\rm eff}$ via $\frac{N_{\rm eff}}{N_{\rm tot}} \equiv (\frac{\sigma_{\rm tot}}{\sigma_{\rm tot,weighted}})^{2}$ we obtain

\begin{equation}
\frac{N_{\rm eff}}{N_{\rm tot}} = \frac{\sum^{i=N_{\rm tot}}_{i=0}\sigma_d^{2}}{\sum^{i=N{\rm tot}}_{i=0} (Z_i\sigma_d)^{2}},
\end{equation}
which we can divide $\sigma_d$ out of to yield
\begin{equation}
\frac{N_{\rm eff}}{N_{\rm tot}} = \frac{N_{\rm tot}}{\sum^{i=N{\rm tot}}_{i=0} Z_i^{2}}.
\end{equation}
We take this into account to obtain our effective number density, $n(z)_{\sigma_z}$, which replaces $N_{\rm eff}$. Instead of $N_{\rm tot}$ samples, we have $N_{\rm tot}$ different weights $Z$ being applied to portions of the sample with different redshift uncertainty, each with a number density $n(\sigma_z)$. We find 
\begin{equation}
n(z)_{\sigma_z} = n_{\rm tot}(z)^2/\sum^{i=N_{\rm tot}}_{i=0} n_i(\sigma_z)Z_i^2(\sigma_z).
\end{equation}
 The redshift uncertainty is simply the mean weighted redshift uncertainty
\begin{equation}
\langle \sigma_{z}\rangle = \sum^{i=N_{\rm tot}}_{i=0} Z_i(\sigma_z)\sigma_{z,i} n_i(\sigma_z)/\sum^{i=N{\rm tot}}_{i=0} Z_i(\sigma_z) n_i(\sigma_z).
\end{equation}

We start with a simple example. We consider a total sample with a number density $2\times 10^{-3}h^3$Mpc$^{-3}$, evenly divided between galaxies with $\sigma_z/(1+z) = 0.03$ and $\sigma_z/(1+z) = 0.05$. We assume a galaxy bias of 1.8 and a redshift of 0.8. For this particular case, we search for the $Z_2/Z_1$ that minimizes the forecasted BAO uncertainty and find that a relative weighting of 0.44 provides the optimal result. This optimal result is, however, only a 3 per cent improvement over the case where only the galaxies with $\sigma_z = 0.03(1+z)$ are used. If both galaxy populations were to be used without any weighting, the weighted combination provides a 2 per cent improvement. If instead, we consider the case where the number density of the galaxies with $\sigma_z = 0.03(1+z)$ is $10^{-4}h^3$Mpc$^{-3}$, we find the relative weighting factor is instead 0.72 and the weighting allows a 0.4 per cent improvement in the constraints. These results are presented in Table \ref{tab:baosigz}, along with a few other examples. 

We find that the relative gains from the optimal weighting procedure vary based on $\sigma_{z1}/\sigma_{z2}$ and the respective number densities, e.g., the relative gain is the same for $\sigma_{z1}=0.02, \sigma_{z2}=0.04$ and $\sigma_{z1}=0.03, \sigma_{z2}=0.06$, though the relative weighting $Z_2/Z_1$ changes. One such example is shown in the top two rows of Table \ref{tab:baosigz}. In Fig. \ref{fig:baosigzw}, we show the percentage improvement achieved by applying optimal weights to each sample as a function of $\sigma_{z1}/\sigma_{z2}$, given different choices for the number density of each sample (defined in this manner, the curves should be symmetric around $\sigma_{z1}/\sigma_{z2}=1$ and we ignore $\sigma_{z1}/\sigma_{z2}<1$). We find at most a 2.5 per cent improvement. The black curve turns over, as at a certain redshift uncertainty the improvement becomes relative to the uncertainty obtained from the sample with lower redshift uncertainty, rather than that of the equally-weighted combination. We find no turn-over for the cases where the number densities are not matched, though we have not tested beyond $\sigma_{z1}/\sigma_{z2}>2$.

\begin{figure}
\includegraphics[width=84mm]{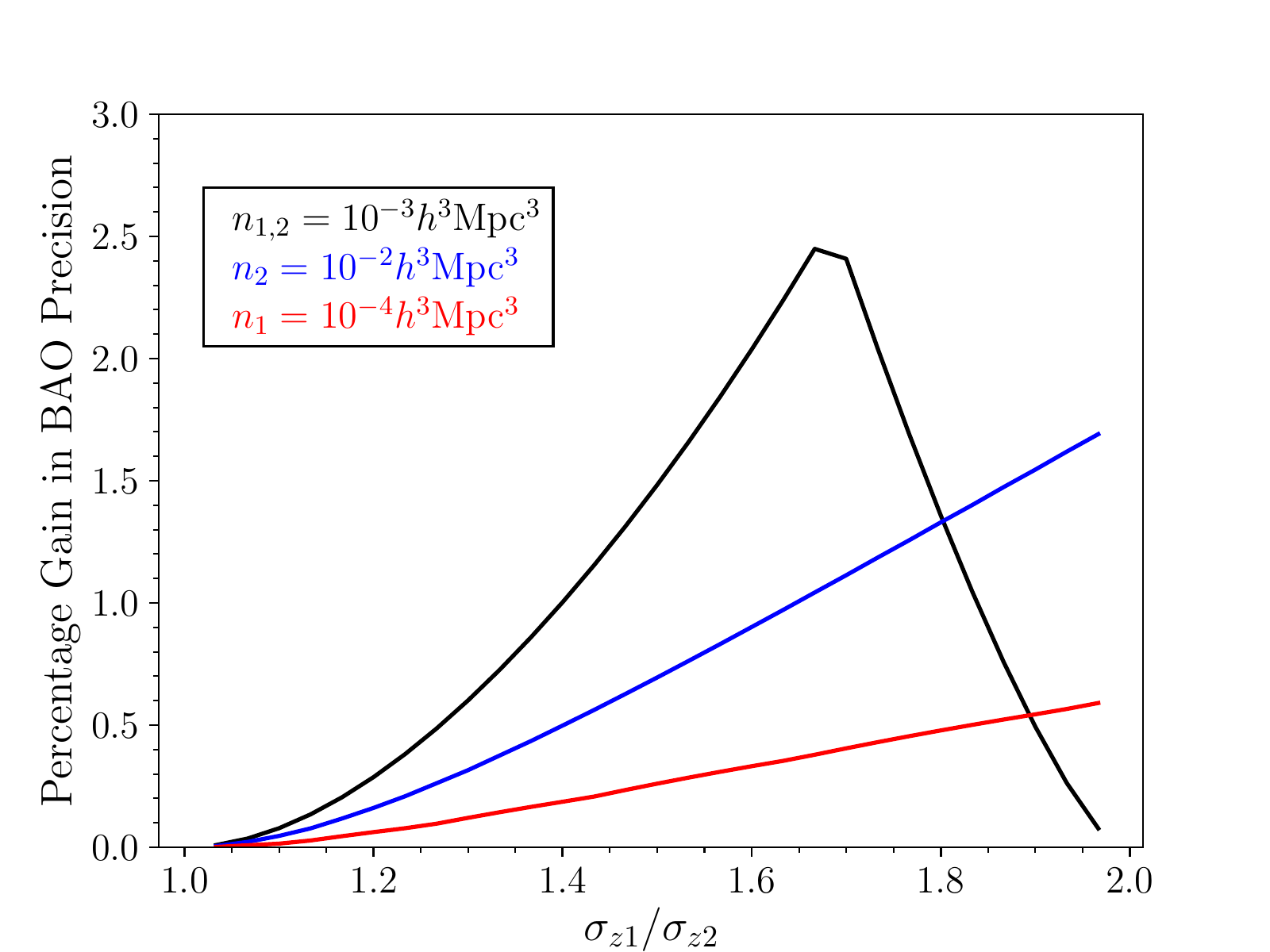}
 \caption{ The percentage improvement obtained in the BAO precision as a function of the ratio between the photometric redshift uncertainties. All cases assume $z=0.8$, $b=1.8$, $n_1, n_2=10^{-3}h^3$Mpc$^{-3}$ unless labeled otherwise. The percentage gain is relative to the optimal choice between either weighting each sample evenly or omitting one sample. }
 \label{fig:baosigzw}
\end{figure}

\begin{table}
\centering
\caption{Given the properties of two samples occupying the same volume, we find the optimal weighting for each and the forecasted uncertainty for this combination divided by the forecasted uncertainty for each individual sample and their evenly-weighted combination. The $\sigma_z$ are divided by $1+z$ and the number densities, $n$, have units $10^{-3}h^3$Mpc$^{-3}$. The optimal relative weighting of the samples is denoted by $Z_2/Z_1$. The quantities $\sigma_{\rm opt}, \sigma_{1}, \sigma_{2}, \sigma_{1+2}$ refer to the Fisher matrix forecast for the BAO uncertainty. At most, we find that weighting for the redshift uncertainty improves the results by 2 per cent (the results in the right-most column). The number density used to obtain $\sigma_{1+2}$ is $n_1 + n_2$ and redshift uncertainty is $\bar{\sigma}_z = 1/2\sum_i \sigma_{z,i}n_i$; these would be the nominal quantities one uses in the case with no weighting.}
\begin{tabular}{lccccccccc}
\hline
\hline
$\sigma_{z,1}$ & $\sigma_{z,2}$ & $\bar{\sigma}_z$ &$n_1$  & $n_2$  & $Z_2/Z_1$ & $\frac{\sigma_{opt}}{\sigma_{1}}$ & $\frac{\sigma_{opt}}{\sigma_{2}}$ & $\frac{\sigma_{opt}}{\sigma_{1+2}}$\\
\hline
0.04 & 0.067 & 0.054 & 1 & 1 & 0.40 & 0.97 & 0.75 & 0.98\\
0.03 & 0.05 & 0.048 & 1 & 10 & 0.44 & 0.85 & 0.96 & 0.99\\
0.03 & 0.05 & 0.043 & 0.5 & 1 & 0.58 & 0.87 & 0.83 & 0.99\\
0.03 & 0.05 & 0.048 & 0.1 & 1 & 0.72 & 0.46 & 0.95 & 0.996\\
0.02 & 0.05 & 0.047 & 0.1 & 1 & 0.61 & 0.56 & 0.94 & 0.99\\
0.02 & 0.03 & 0.027 & 0.4 & 1 & 0.65 & 0.81 & 0.87 & 0.99\\
\hline
\label{tab:baosigz}
\end{tabular}
\end{table}

The discussion above provides a framework that one could expand upon to obtain an optimal set of weights based on the redshift uncertainty. For the cases considered in Table \ref{tab:baosigz} and Fig. \ref{fig:baosigzw}, we find that applying optimal weights provides at most a 2.5 per cent improvement over what would be considered the optimal sample selection. This suggests that for the number densities and redshift uncertainties achievable by multi-band imaging surveys such as DES, weighting by individual redshift uncertainties is unlikely to make a substantial impact on the recovered results. These results therefore validate our approach to treat galaxy samples as having a single redshift uncertainty, equal to the mean uncertainty of a given population of galaxies. Our simple tests reveal at most a 2.5 per cent improvement over this approach (assuming one removes galaxies with redshift uncertainties that degrade the ensemble constraints).

\section{Measuring Clustering in Mock Galaxy Samples}
\label{sec:mocks}

In order to test and validate the analytic results we presented in the previous Section, we conduct clustering analyses on mock galaxy catalogs (mocks). In this section, we first describe how we created mocks that simulate DES Y1 data and then describe the clustering we measure for those mocks.

\subsection{HALOGEN Mocks} 
\label{sec:halogen}

We test and validate our analytical results using two sets of mocks, produced to have similar properties to the DES Y1 BAO sample: `Square' mocks and `Y1' mocks.
We generate the mocks using HALOGEN \citep{halogen}, a method based on an exponential bias model applied to a 2nd-order Lagrangian perturbation theory (2LPT) density field. As a first step in their production, HALOGEN dark matter halo catalogs are produced such that the halo clustering and velocity distributions match those of the MICE \citep{CrocceMICE,FosalbaMICE,FosalbaMICE2,CarreteroMICE} $N$-Body simulation as a function of both mass and redshift. 

We produce lightcone catalogs by super-imposing redshift shells (each of them obtained from a cubic snapshot of length 3072 $h^{-1}{\rm Mpc}$) of width $\Delta_z=0.05$ in the interval $0.6<z<1$, and two additional slices from snapshots $z=0.55$ and $z=1.05$ to complete the boundaries of the lightcone (and for the Y1 mocks another two at $z=0.3$ and $1.3$). 
From each set of snapshots we produce eight independent halo lightcone catalogs to a depth of $z=1.42$
and mass resolution of  $M=2.5\cdot10^{12}M_{\odot}/h$. 

Our first set of mocks, denoted `Square', simulate early estimates of the properties of the DES Y1 BAO sample using 504 halo catalogs. This data sample occupies 1420 deg$^2$ with a number density that decreased from $3\times10^{-3}h^3$Mpc$^{-3}$ at $z=0.6$ to $8\times10^{-4}h^3$Mpc$^{-3}$ at $z=1$ and redshift uncertainty $\sigma_z = 0.029(1+z)$ (these redshift uncertainties were later found to be optimistic). At the resolution of our halo catalogs, a maximum number density of $2\times10^{-3}h^3$Mpc$^{-3}$ is achievable at $z=0.6$. Thus we increased the area in the footprint of our Square mock samples to $A_{\rm mock}=1779$ deg$^2$. We used the \cite{SeoEis07} Fisher matrix predictions to determine that this is the area required to be able to match the estimate of BAO signal to noise for the mocks and (early estimates of) the data as a function of redshift. Given this expanded area, the signal to noise as a function of redshift was matched by adjusting the number density in intervals of $\Delta_z=0.05$, obtaining the density displayed in Fig. \ref{fig:nz}. Halos were then selected by mass, yielding the bias evolution represented in that Fig. \ref{fig:bias}. 
For simplicity, we use a \textit{square} mask, selecting galaxies to be at the center of halos in with both $\phi$ and ${\rm cos}(\theta)$ in the interval $[0,0.7361]$ (with $\theta,\phi$ representing the latitudinal and longitudinal coordinates of an arbitrarily sized sphere). Given its geometry, we refer to this mock sample as the `Square' sample.
The process described above gave us mock galaxy catalogs with angular coordinates and cosmological redshifts. We added the effects of RSD by correcting the redshifts based on the peculiar velocity of each halo. We then added the redshift uncertainty by sampling a Gaussian centered at this redshift of width $\sigma_z=0.029(1+z)$. 

Our second set of mocks resembles more closely the true nature of the final DES Y1 BAO sample (Crocce et al., in prep.),
 we refer to it as the `Y1' sample. It matches the footprint (with an area of 1420 deg$^2$), number density (evolving from $1.4\times10^{-3}h^3$Mpc$^{-3}$ at $z=0.6$ to $4\times10^{-4}h^3$Mpc$^{-3}$ at $z=1$), angular clustering, and redshift uncertainty (evolving from $0.028(1+z)$ to $0.05(1+z)$). The number density and bias of these mocks are also shown in Fig. \ref{fig:nz} and \ref{fig:bias}. Other novelties in these mocks include the non-Gaussian modelling of the redshift uncertainty and a redshift-evolving Halo Occupation Distribution galaxy model, both fit to the DES Y1 data. In particular, to match our estimates of the DES Y1 $dN/dz$ as a function of photometric redshift, a non-Gaussian PDF was used to assign the photometric redshift of a galaxy given its true redshift. Full details of the mocks and the data are found, respectively, in 
Avila et al. (in prep.) and Crocce et al. (in prep.).
   We analyze 1200 mocks for this Y1 sample.

\begin{figure}
\includegraphics[width=84mm]{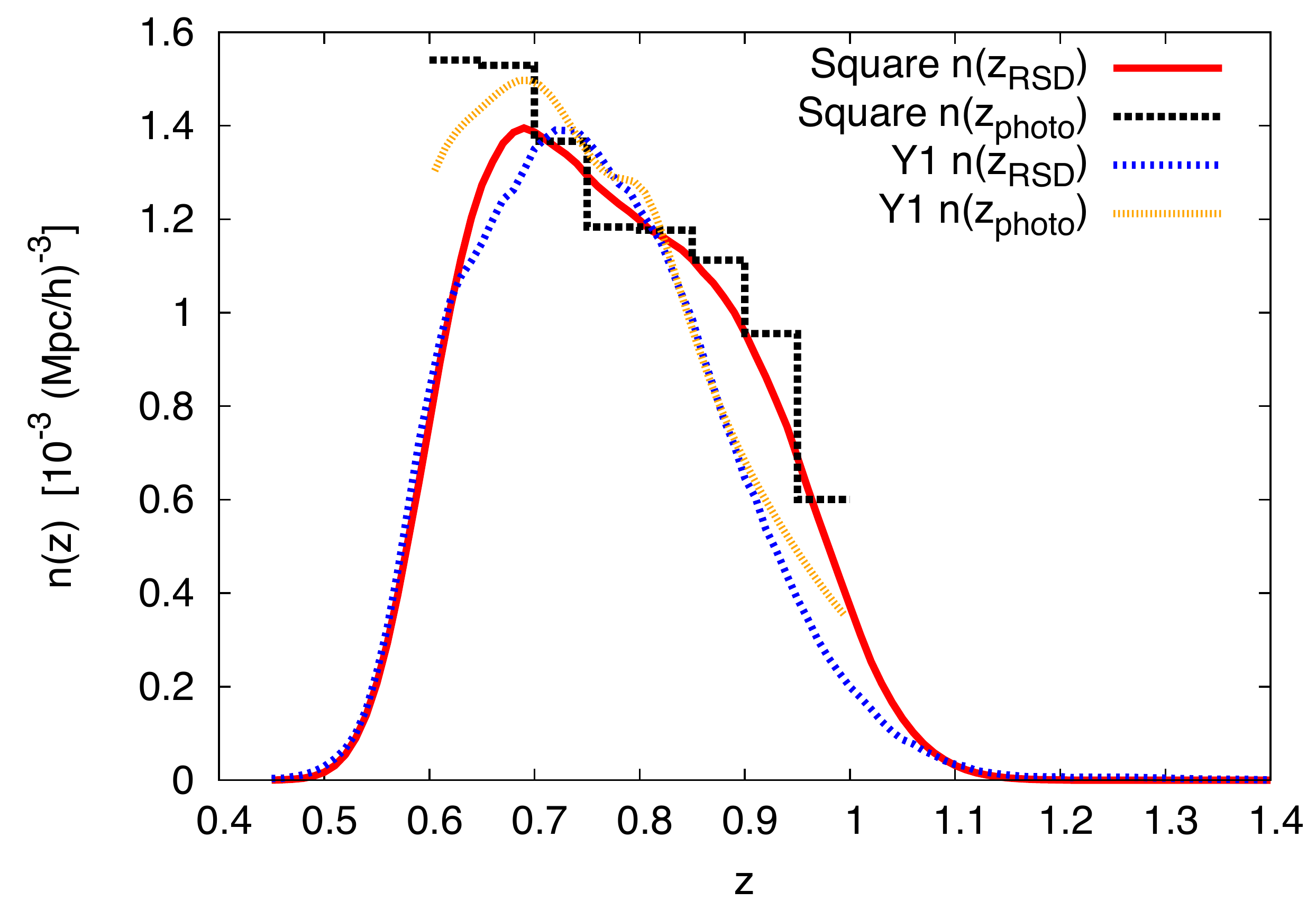}
  \caption{Number density of objects as a function of redshift before and after the application of the photo-z modeling, for both the 504 `Square' and 1200 `Y1' halogen mocks we use. }
  \label{fig:nz}
\end{figure}

\begin{figure}
\includegraphics[width=84mm]{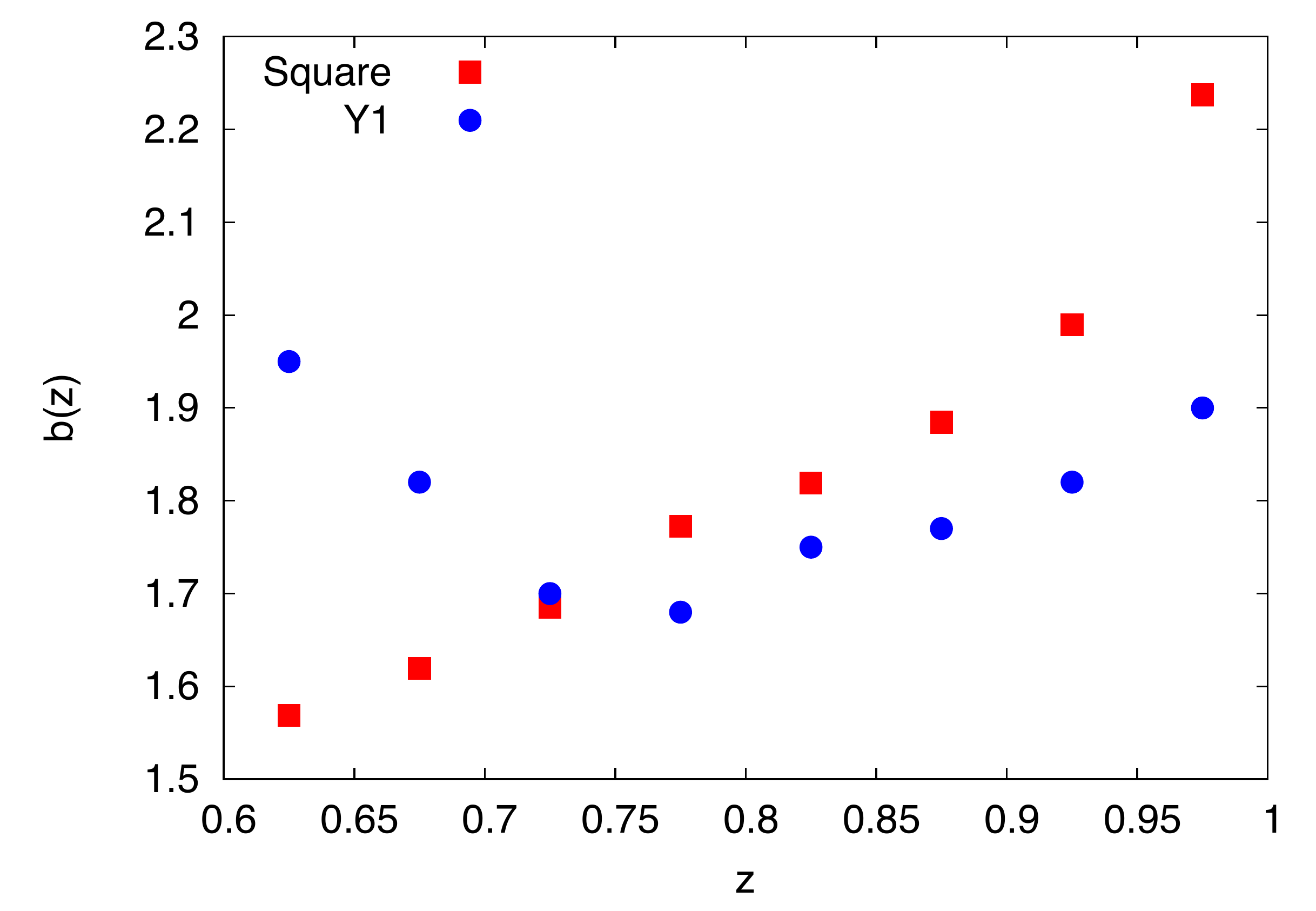}
 \caption{The bias as a function of redshift, for the 504 `Square' and 1200 `Y1' halogen mocks we use. The Y1 mocks are a closer match to the DES Y1 data, but the Square mocks use a simpler methodology and are thus useful for aiding theoretical understanding.}
 \label{fig:bias}
\end{figure}

\subsection{Clustering Measurements}
\label{sec:mockclus}
We measure the correlation functions of our mock galaxy samples, $\xi$, by converting angles and redshifts to physical distances; this gives us a three dimensional map in `photometric redshift space'. In this space, the apparent transverse and radial separations, $s_{\perp}$ and $s_{||}$, are calculated for pairs of galaxies and/between a synthetic random catalog matching the angular and redshift selections of the mocks. Given (normalized) pair counts, we calculate $\xi_{\rm phot}(s_{\perp},s_{||})$ via \cite{LS}
\begin{equation}
\xi_{\rm phot}(s_{\perp},s_{||}) =\frac{DD(s_{\perp},s_{||})-2DR(s_{\perp},s_{||})+RR(s_{\perp},s_{||})}{RR(s_{\perp},s_{||})}, 
\label{eq:xicalc}
\end{equation}
where $D$ represents the galaxy sample and $R$ represents the uniform random sample that simulates the selection function of the galaxies. $DD(s_{\perp},s_{||})$ thus represent the number of pairs of galaxies with separation $s_{\perp}$ and $s_{||}$, within some bin tolerance. We use bins of 1$h^{-1}$Mpc . These are narrow enough to allow us to test many treatments with respect to $\mu_{\rm obs} = s_{||}/\sqrt{s^2_{\perp}+s^2_{||}}$.

Our use of the clustering statistic $\xi(s_{\perp})$ within some range of $\mu_{\rm obs}$ is similar to the commonly used $w_p(r_p)$ statistic. We are using the coordinates $s_{\perp},s_{||}$ to have the same meaning as $r_p,\pi$ in, e.g., \cite{Zehavi11}. Using our convention, their equation 3 is
\begin{equation}
w_p(s_{\perp}) = 2 \int_0^{\infty}{\rm d}s_{||}\xi(s_{\perp},s_{||}).
\end{equation}
In practice, studies have employed some maximum $s_{||}$ value. Our approach is different in that we consider some window, $W(\mu_{\rm obs})$
\begin{equation}
\xi_{\rm phot, F}(s_{\perp}) =  \int_0^{1}{\rm d}\mu_{\rm obs} W(\mu_{\rm obs})\xi_{\rm phot}(s_{\perp},s_{||}),
\end{equation}
normalized such that $\int_0^1 {\rm d}\mu_{\rm obs} W(\mu_{\rm obs}) = 1$. The main practical difference between our approach and that of $w_p(s_{\perp})$ is that our results will be binned in terms of $\mu_{\rm obs}$ and any cut on $\mu_{\rm obs}$ results in a varying maximum $s_{||}$ as a function of $s_{\perp}$.

In Eq. \ref{eq:wfkpbz}, we defined a weight, $w_{\rm FKP}$, to be applied as a function of redshift, given a calculation of  $n_{\rm eff}(z)$ and $k_{\rm eff}$. This weight is meant to properly account for the changing signal to noise in each redshift shell. We apply these weights to the galaxies and random points and thus each paircount is weighted by the multiplication of the two weights representing each pair. 

Our $w_{FKP}(z)$ weights were defined as follows. Based on the sample characteristics presented in Section \ref{sec:halogen}, we find that $k_{\rm eff}$ is within 10 per cent of $0.12h$Mpc$^{-1}$ for each sample of mocks, independent of redshift. We thus use a value $P_0 = 4500h^{-3}$Mpc$^{3}$, which is close to the linear redshift zero matter power spectrum at $k = 0.12h$Mpc$^{-1}$ for our fiducial (MICE) cosmology. For the Square mocks, $n_{\rm eff}$ evolves from $1.8\times10^{-4}h^{3}$Mpc$^{-3}$ to $6.9\times10^{-5}h^{3}$Mpc$^{-3}$. Using Eq. \ref{eq:wfkpbz}, we find the $w_{FKP}(z)$ evolves from 0.82 to 1 for this sample (normalizing so that the maximum weight value is 1). For the Y1 mocks, we follow a similar process and find the weights evolve from 0.71 to 1. We therefore expect the use of these weights to have a minor effect on our results, as the arguments at the end of Section \ref{sec:zweight} suggest only a 5 per cent gain in precision when the weights differ by a factor of 2.

\begin{figure}
\includegraphics[width=84mm]{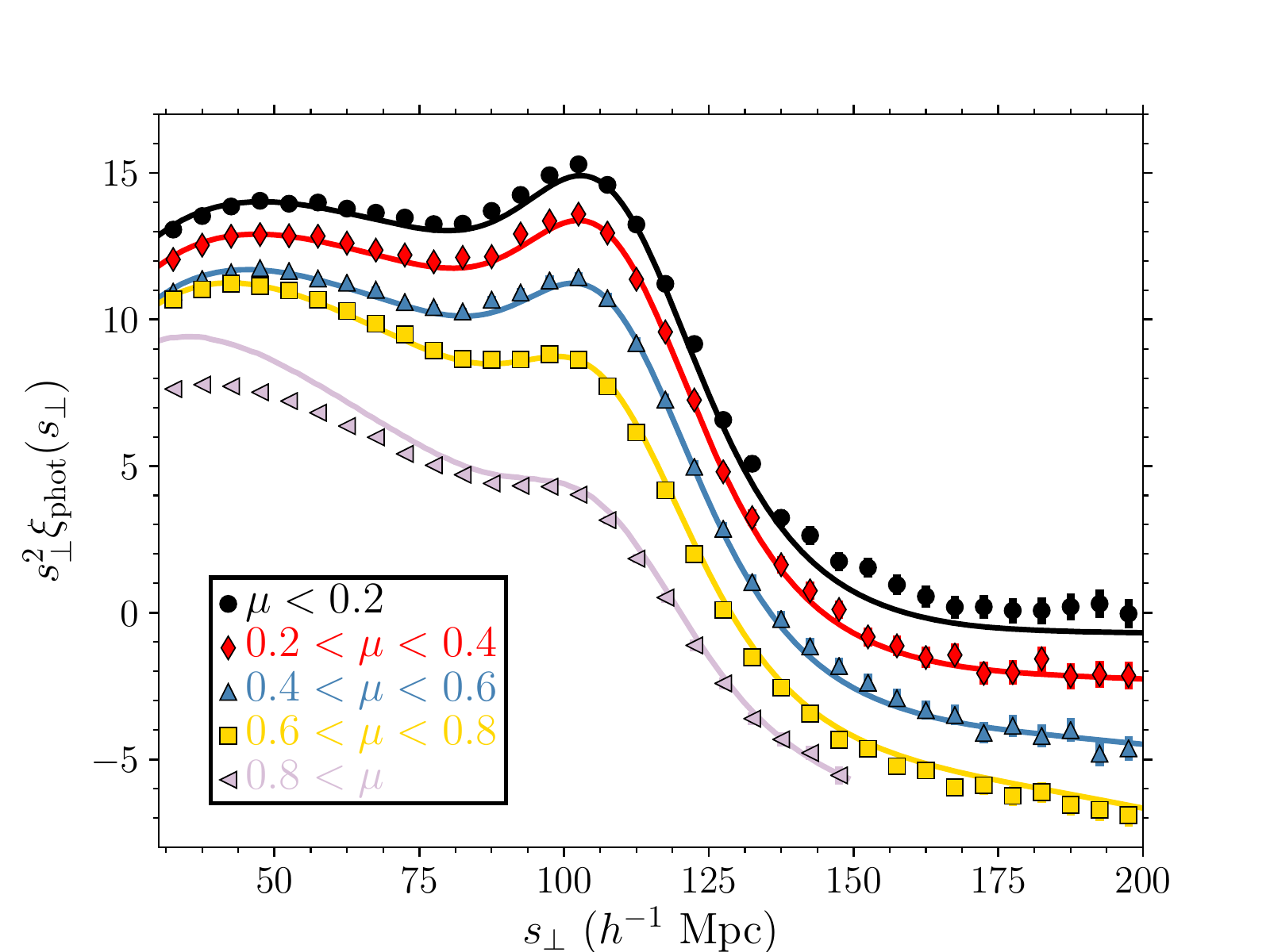}
  \caption{The predicted $\xi_{\rm phot}(s_{\perp})$ signal (solid curves; obtained via Eqs. \ref{eq:xip}-\ref{eq:xiphot}) in bins of observed $\mu$ {\bf (labeled $\mu$ in the legend)}, the cosine of the angle to the line of sight, compared to the mean of 504 `Square' mock realizations. The mock points all have $5\times10^{-5}$ subtracted from them, before multiplying by $s_{\perp}^2$. The error-bars are the standard deviation of the mocks samples divided by $\sqrt{504}$. The model curves assume a sample centered on $z=0.8$ with $\sigma_z/(1+z) = 0.029$, and a bias of 1.76, as is approximately the case for the mock samples. The red curve and points are shifted vertically by 1 and the black curve and points by 2, for legibility. }
  \label{fig:modcompmock}
\end{figure}

Fig. \ref{fig:modcompmock} displays the mean of the clustering in the Square mock samples, divided into 0.2 thick bins in $\mu_{\rm obs}$ and binned in terms of the transverse separation $s_{\perp}$. We have subtracted $5\times10^{-5}$ from each set of points and then multiplied by $s_{\perp}^2$. The error bars represent the standard deviation across the 504 mock realizations, divided by $\sqrt{504}$; i.e., this is the ensemble uncertainty. The curves represent the model described in previous sections, with a bias of 1.76 (consistent with the $b(z)$ given in Fig. \ref{fig:bias} and our redshift weighting). The location of the BAO feature remains nearly constant as a function of $\mu_{\rm obs}$; this is because at high $\mu_{\rm obs}$, the vast majority of the information is from pairs with {\it true} $\mu$ values that are close to zero. While the BAO feature stays nearly constant (see Fig. \ref{fig:BAOsig6pan}), the shape changes considerably with $\mu_{\rm obs}$, such that the overall amplitude is decreased with increasing $\mu_{\rm obs}$. For $\mu_{\rm obs} < 0.8$ and $s_{\perp} < 140 h^{-1}$Mpc, the model curves are a good match to the mock measurements, given the $5\times10^{-5}$ offset.\footnote{We are unable to determine the reason for the mismatch between the model and mock results for $\mu_{\rm obs} > 0.8, s_{\perp} < 80h^{-1}$Mpc, but given the lack of BAO signal for $\mu_{\rm obs} > 0.8$ this does not impact strongly impact our results.} Our pair-counts were calculated for $r_{||} < 200 h^{-1}$Mpc and thus there are no results for $\mu_{\rm obs} > 0.8, s_{\perp} > 150h^{-1}$Mpc. 

The match between the Y1 mocks and our theoretical calculation is not displayed, but is similarly good. The agreement is achieved despite the fact that the Y1 mocks do not assume Gaussian redshift uncertainties, while our modeling does. Further details can be found in Avila et al. (in prep.). %
For the Y1 mocks, we will compare results to those obtained from the angular correlation function, $w(\theta)$. We calculate $w(\theta)$ as in Eq. \ref{eq:xicalc}, except that we bin paircounts by their angular separation, $\theta$. When doing so, we divide the measurements into four redshift bins between $0.6 < z_{\rm phot} < 1.0$, each with thickness $\Delta_{z} = 0.1$.

\section{BAO Information in Mock Samples}
\label{sec:BAOmock}

In this section, we compare measurements of the BAO in mock samples for different $\mu_{\rm obs}$ windows and test the extent to which the $\int d\mu_{\rm obs} W_{\rm opt}(\mu_{\rm obs})\xi_{\rm phot}(s_{\perp},\mu_{\rm obs})$ described in Section \ref{sec:BAOsn} is indeed the optimal estimator for BAO information.

In order to extract the BAO information, we fit to a model (c.f., \citealt{Xu12,alph,Ross16})
\begin{equation}
\xi_{F, {\rm mod}}(s_{\perp}) = B_F\xi_{F, {\rm phot}}(s_{\perp}\alpha) + A_{F}(s_{\perp})  
\label{eq:xiFmod}
\end{equation}
where $\xi_{F, {\rm phot}}$ represents the use of a general $\mu_{\rm obs}$ window (i.e., $\xi_{F, {\rm phot}} = \int d\mu_{\rm obs} W_{F}(\mu_{\rm obs})\xi_{\rm phot}(s_{\perp},\mu_{\rm obs})$) and $A_F(s_{\perp}) = a_{F,1}/s_{\perp}^2+a_{F,2}/s_{\perp}+a_{F,3}$. In each case, the parameter $B_F$ essentially sets the size of the BAO feature in the template. We apply a Gaussian prior of width ${\rm log}(B_F) = 0.4$ around the best-fit $B_F$ in the range $50 < s < 80h^{-1}$Mpc with $A_F = 0$. Likelihoods for $\alpha$ are obtained by finding the minimum $\chi^2$ on a grid of $\alpha$ between 0.8 and 1.2, with spacing 0.001, when marginalizing over the other model parameters. This is a close match to the methods used in BOSS (see, e.g., \citealt{Xu12,alph,Ross16} and references therein). The $\chi^2$ are determined in the standard manner, with the covariance determined from the sample of mocks being tested, i.e.,
\begin{equation}
C(s_1,s_2) = \frac{1}{N_{\rm mock}-1}\sum_i\left([\xi(s_1)-\langle\xi(s_1)\rangle][\xi(s_2)-\langle\xi(s_2)\rangle]\right)
\end{equation}

We start by using the 504 Square mocks in order to explore how the signal to noise varies as a function of $\mu_{\rm obs}$. While we have fewer Square mocks compared to the Y1 mocks, they have better expected precision and a simple constant $\sigma_z = 0.029(1+z)$, making them better suited for sub-dividing and comparing to theoretical expectation. We use $\xi_{\rm phot}(s_{\perp})$ measurements averaged in bins of 0.2 thickness in $\mu_{\rm obs}$ to measure the BAO location and its uncertainty. The results are presented in the top rows of Table \ref{tab:baomock}. The uncertainty we recover from the mean of the mock samples (denoted `mean' in the table) increases slightly with increasing $\mu_{\rm obs}$, from 0.039 to 0.041 for $\mu_{\rm obs} < 0.8$ and then sharply to 0.051 for $\mu_{\rm obs}>0.8$. These results for the uncertainty from the mean of the mocks are displayed with yellow stars in Fig. \ref{fig:relinfo}. One can see that the distribution of information with apparent $\mu_{\rm obs}$ is close to our Fisher matrix analysis, except that we find a more gentle decrease in the information recovered as $\mu_{\rm obs}$ increased than compared to the prediction. 

The $\mu_{\rm obs} > 0.8$ clustering contains significantly less BAO information than the rest of the $\mu_{\rm obs}$ range. Given the lack of constraining power and the increased difficulty in its modeling, we ignore the $\mu_{\rm obs} > 0.8$ data from further comparisons. Given the covariance between data at different $\mu_{\rm obs}$ (described later in this section), we expect this has minimal impact on the BAO results we recover from the mocks.

If we instead look at the mean uncertainty for mocks where we find a 1$\sigma$ bound in region $0.8 < \alpha < 1.2$, denoted `samples' in the table, we find similar results. The mean uncertainty increases from 0.039 to 0.041 with increasing $\mu_{\rm obs}$. We also find the standard deviation of the recovered maximum likelihood $\alpha$ values increases and is in the range 0.042 to 0.044. Correspondingly, the fraction of mocks with 1$\sigma$ bounds also decreases slightly, from 0.98 to 0.96. Including only mocks with 2$\sigma$ bounds within the region $0.8 < \alpha < 1.2$, the mean uncertainties are, naturally, smaller by $\sim$ 10 per cent. For this case, the standard deviations are closer to the mean uncertainty; this is consistent with the likelihoods for the better-constrained realizations being more Gaussian. Our results from splitting the clustering by  $\mu_{\rm obs}$ consistently show a slight decrease in BAO information content with increasing $\mu_{\rm obs}$ in the range $0 < \mu_{\rm obs} < 0.8$.

The results are improved when we combine all data with $\mu_{\rm obs} < 0.8$. These results for the Square mocks are found in the middle rows of Table \ref{tab:baomock}. Given the similarity of the BAO signal shown in the bottom-right panel of Fig. \ref{fig:BAOsig6pan}, we expect such a compression to preserve the strength of the BAO signal. The uncertainty on the mean of the mock samples decreases to 0.038 and we find the mean of the uncertainty determined for each individual sample is the same (when considering the 1$\sigma$ threshold). The standard deviation is 0.042 and the fraction of mocks with a 1$\sigma$ bound matches that for $\mu_{\rm obs}<0.2$ (0.98). The dotted line in Fig. \ref{fig:relinfo} displays how we expect BAO information to be distributed in apparent $\mu_{\rm obs}$ and approximately matches what we find for the mocks divided in $\mu_{\rm obs}$ bins. This suggests that we should use that curve as a relative weight to apply as a function of $\mu_{\rm obs}$. We do so and find negligible improvement. For the mean of the mocks, we find a 0.2 per cent improvement in the recovered uncertainty (which is too small to be observable in Table \ref{tab:baomock}). For the mock realizations, the biggest change is a 1 per cent improvement in the standard deviation obtained when limiting to those cases with 1$\sigma$ bounds within $0.8 < \alpha < 1.2$.

The precision of the results for $\mu_{\rm obs} < 0.8$ are just slightly better than those obtained for more thin slices in $\mu_{\rm obs}$. This is due to fact that the results for different $\mu_{\rm obs}$ slices are highly correlated (as implied by the fact that the information is sharply peaked at low {\it true} $\mu$). The correlation coefficients vary from 0.77 to 0.7 (these can be compared to correlation coefficients of $\sim$0.15 for mocks simulating a spectroscopic redshift sample in \citealt{Ross152D}). One can construct a covariance matrix for the four $\mu_{\rm obs}$ bins and take the sum of its inverse to obtain a standard deviation from the combination of the $\mu_{\rm obs}$ bins. We use mock realizations with 2$\sigma$ bounds and find 0.034, which can be compared to 0.036 for our $\mu_{\rm obs} < 0.8$ results. This suggests that $\xi(s_{\perp})$ with $\mu_{\rm obs} < 0.8$ is close to an optimal compression of the BAO information distributed in the four $\mu_{\rm obs}$ bins, but does suffer $\sim$10 per cent information loss ($[0.036/0.034]^2 = 1.12$). 

For the Y1 mocks, we focus on $\mu_{\rm obs} < 0.8$ and weight evenly. The redshift uncertainties are greater for the Y1 mocks, suggesting the information should be spread more evenly in observed $\mu_{\rm obs}$ than for the Square mocks. We use 1200 Y1 realizations to define the covariance matrix. The uncertainty obtained for the mean of the mocks is just greater than five per cent and this is broadly consistent with the results obtained from the individual mock realizations. More than 90(70) per cent of the realizations have a 1(2)$\sigma$ bound within $0.8 < \alpha < 1.2$. This suggests there is a good likelihood of obtaining a robust 5 per cent angular diameter distance measurement using BAO in the DES Y1 data. 

For Y1 mocks, we can compare to $w(\theta)$ results (we do not have enough Square mocks for a robust covariance matrix for $w(\theta)$). We find a 13 per cent improvement in the uncertainty obtained on the mean when using $\xi_{\rm phot}(s_{\perp})$. Here, for $w(\theta)$, we are applying the same BAO model and simply swapping $\xi_{\rm phot}$ \&$s_{\perp}$ for $w$ \& $\theta$. A more detailed comparison of methodologies will be presented with the DES collaboration analysis of the Y1 BAO signal. 
This study will include further tests to optimize each method. In principle, the precision of $w(\theta)$ results should converge to the $\xi_{\rm phot}(s_{\perp})$ results as the redshift bin size is narrowed and all cross-correlations between redshift bins are included. The clear advantage of $\xi_{\rm phot}(s_{\perp})$ is the smaller size of the data vector, which reduces the noise bias in the inverse covariance matrix \citep{Hartlap07,Dod13,Per14}.

\begin{table}
\centering
\caption{ 
Statistics for BAO fits on mocks. $\langle\alpha\rangle$ is either the BAO dilation-scale measured from the correlation function averaged over all of the mocks (denoted `mean'), or the mean of the set of dilation-scales recovered from mocks with $>1\sigma$ ($>2\sigma$) BAO detections (denoted `samples'). $\langle\sigma\rangle$ is the same for the uncertainty obtained from $\Delta\chi^2=1$ region. $S$ is the standard deviation of the $\alpha$ recovered from the mock realizations with $>1\sigma$  ($>2\sigma$) BAO detections and $f(N_{\rm det})$ is the fraction of realizations with $>1\sigma$ ($>2\sigma$) detections. 
The label $w_{\mu}$ denotes that the results have been weighted in $\mu_{\rm obs}$ following the dotted line in Fig. \ref{fig:relinfo}.}
\begin{tabular}{lcccc}
\hline
\hline
case & $\langle\alpha\rangle$ & $\langle\sigma\rangle$ & $S$ & $f(N_{\rm det})$\\
\hline
Square mocks, $\mu_{\rm obs}$ bins:\\
mean, $\mu_{\rm obs} < 0.2$ & 1.007 & 0.039 & - & -\\
\ samples, $\mu_{\rm obs}< 0.2$, $1\sigma$ & 1.007 & 0.039 & 0.042 & 0.976\\
\ samples, $\mu_{\rm obs}< 0.2$, $2\sigma$ & 1.006 & 0.037 & 0.037 & 0.885\\
mean, $0.2 < \mu_{\rm obs} < 0.4$, $1\sigma$ & 1.006 & 0.040 & - & -\\
\ samples, $0.2 < \mu_{\rm obs} < 0.4$, $1\sigma$ & 1.005 & 0.039 & 0.042 & 0.976\\
\ samples, $0.2 < \mu_{\rm obs} < 0.4$ , $2\sigma$ & 1.004 & 0.038 & 0.038 & 0.901\\
mean, $0.4 < \mu_{\rm obs} < 0.6$ & 1.005 & 0.041 & - & -\\
\ samples, $0.4 < \mu_{\rm obs} < 0.6$, $1\sigma$ & 1.005 & 0.041 & 0.044 & 0.966\\
\ samples, $0.4 < \mu_{\rm obs} < 0.6$, $2\sigma$ & 1.003 & 0.038 & 0.038 & 0.855\\
mean, $0.6 < \mu_{\rm obs} < 0.8$ & 1.005 & 0.041 & - & -\\
\ samples, $0.6 < \mu_{\rm obs} < 0.8$, $1\sigma$ & 1.004 & 0.041 & 0.044 & 0.958\\
\ samples, $0.6 < \mu_{\rm obs} < 0.8$, $2\sigma$& 1.003 & 0.037 & 0.039 & 0.813\\
mean, $\mu_{\rm obs} > 0.8$ & 0.993 & 0.051 & - & -\\
\hline
Square mocks, $\mu_{\rm obs} < 0.8$:\\
mean, $w_{\mu}$ & 1.008 & 0.038 & - & -\\
\ samples, $w_{\mu}$, $1\sigma$ & 1.008 & 0.038 & 0.041 & 0.976\\
\ samples, $w_{\mu}$, $2\sigma$ & 1.007 & 0.035 & 0.036 & 0.889\\

mean & 1.008 & 0.038 & - & -\\
\ samples, $1\sigma$ & 1.008 & 0.038 & 0.042 & 0.978\\
\ samples, $2\sigma$ & 1.007 & 0.035 & 0.036 & 0.883\\
\hline
Y1 mocks, $\mu_{\rm obs} < 0.8$:\\
mean & 1.005 & 0.053 & - & -\\
\ samples, $1\sigma$ & 1.003 & 0.047 & 0.053 & 0.910\\
\ samples, $2\sigma$ & 1.000 & 0.042 & 0.044 & 0.706\\
mean $w(\theta)$, $\Delta z=0.1$ & 1.009 & 0.061 & - & - \\
\hline

\label{tab:baomock}
\end{tabular}
\end{table}

The mean $\alpha$ values for the Square mocks are slightly biased compared to the expectation of $\alpha =1$, though the Y1 mocks are less biased. Some of this bias likely comes from non-linear structure, which is expected produce shifts less than 0.005 (c.f., \citealt{Seo08,Pad12}). The biases we find for the Square mocks increase as $\mu_{\rm obs}$ gets smaller, which is the opposite as we would naively expect, given the modeling is more simple for $\mu = 0$. Accounting for the shift expected from non-linear structure growth, these biases are $\sim 0.1\sigma$ and thus not especially concerning for DES Y1. However, future data sets will provide considerably more precise results and will thus require improvements in the modeling if the biases are similar to what we find for the Square mocks. For the Y1 mocks, we observe that the fact that we have modeled the sample assuming Gaussian redshift uncertainties does not significantly bias our results, though it is possible this causes the estimated uncertainties to be smaller than the recovered standard deviations. This will be studied further in a future publication describing the DES Y1 BAO analysis.

We obtain a BAO uncertainty of 0.038 from the Square mocks and 0.053 for the Y1 mocks with $\mu_{\rm obs} < 0.8$. These can be compared to Fisher matrix predictions of 0.031 and 0.038. Our fiducial cosmology closely matches that used to define the \cite{SeoEis07} Fisher matrix forecasts. As explained in \cite{SeoEis07}, this Fisher matrix prediction neglects to include additional BAO smearing due to projection effects, whose effect is described in \cite{SeoEis03}. Essentially, projection effects cause the BAO feature at different redshift to be combined and thus diluted. Emulating \cite{SeoEis03}, we estimate the projection effect should degrade our uncertainties by an additional $\sim$ 10 per cent and we obtain forecasts of 0.033 for the Square mocks and 0.042 for the Y1 mocks. The values we recover from the mean of the mocks are affected somewhat by the likelihoods being non-Gaussian. We can account for this by multiplying the covariance matrix by an arbitrary factor, $f$, small enough so that the recovered $\sigma/\sqrt{f}$ is constant as $f$ is decreased (since for Gaussian likelihoods $\sigma(f)=\sqrt{f}\sigma(f=1)$), i.e., we find the region where the likelihood is Gaussian and scale this appropriately. Doing so, we find a `Gaussian' uncertainty of 0.047 for the mean of the Y1 mocks and no change for the Square mocks. This suggests we are recovering 80 per cent of the available BAO information in both cases\footnote{Determining relative information achieved as compared to that possible as $(\sigma/\sigma_{\rm Fisher})^{-2}$.}. Potential reasons we have not recovered the full 100 per cent include: our compression of the $\mu_{\rm obs} < 0.8$ BAO signal into one data vector, our cut at $\mu_{\rm obs} < 0.8$, our redshift weights being sub-optimal, and the fact that we are treating the redshift errors as Gaussian in the forecasts. 

The results presented in this section validate our modeling of the BAO information with respect to the line of sight. As expected for redshift uncertainty greater than $0.02(1+z)$, the information and the signal itself is found to be nearly constant in $\mu_{\rm obs}$, allowing one to combine the information at $\mu_{\rm obs}<0.8$ into a single $\xi(s_{\perp})$ measurement and extract nearly all of the available BAO information.

\section{Discussion}
\label{sec:disc}

We have presented a concise analysis of the factors that affect the achievable precision of BAO distance scale measurements made using photometric redshift samples. We use this information in order to determine an optimal clustering estimator for BAO measurements to be applied to DES Y1 data. 

Our analytic work comprised of three pieces:
\begin{itemize}
\item{
We first investigated the form of the BAO signal, in configuration space, by subtracting a smooth `no BAO' model from the nominal model, accounting for redshift uncertainty. We found that the BAO feature is nearly constant as a function of observed $\mu_{\rm obs}$ when plotted against the transverse separation, $s_{\perp}$. See Fig. \ref{fig:BAOsig6pan}.}
\item{
We then utilized Fisher matrix predictions to understand the distribution of BAO with respect to the true and observed LOS. Using the cosine of the angle to the LOS, $\mu$, we found that the information is confined primarily to nearly transverse clustering modes ($\mu=0$). For a redshift uncertainty of $\sigma_z = 0.03(1+z)$, there is more than 100 times as much $D_A(z)$ information as $H(z)$ information. However, when we consider the distribution of information in terms of the observed $\mu_{\rm obs}$, we find it is nearly evenly distributed up to $\mu_{\rm obs}\sim 0.8$. The redshift uncertainty causes true $\mu=0$ information to be distributed over all $\mu_{\rm obs}$. This explains the results we obtained for the observed signal. See Fig. \ref{fig:relinfo}.}
\item{
Finally, we derived weights to apply to galaxies based on the evolution in redshift uncertainty, number density, and bias as a function of redshift. This weight is given by Eq. \ref{eq:wfkpbz}. Further, we learned that at a given redshift, weighting galaxies based on their particular redshift uncertainty provides minimal gains (see Table \ref{tab:baosigz})}.
\end{itemize}

The above findings lead to the following recommendations for obtaining optimal BAO measurements from a photometric redshift sample:
\begin{itemize}
\item{Given a parent sample of galaxy magnitudes, redshifts, and estimates of the redshift uncertainty, one should optimize the sample based on optimizing the Fisher forecast for the BAO uncertainty when applying color and magnitude cuts to change the number density and mean redshift of the sample. This matches the approach used for the DES Y1 BAO sample selection (Crocce et al. in prep.). 
Our results in Section \ref{sec:zpweight} suggest that once selecting a sample based on these criteria, the treatment of variations in the redshift uncertainty within the sample should have negligible impact.} 
\item{Assuming the redshift uncertainty is $\geq 0.02(1+z)$, measurements of the clustering with respect to the transverse separation should provide nearly optimal BAO constraints. Thus, one should measure $\xi_{\rm phot}(s_{\perp},\mu_{\rm obs})$ and use Eq. \ref{eq:wfkpbz} to weight each galaxy as function of redshift.}
\item{One should then compress this information into the clustering estimator $\xi_{\rm opt}(s_{\perp}) = \int d\mu_{\rm obs} W_{\rm opt}(\mu_{\rm obs})\xi_{\rm phot}(s_{\perp},\mu_{\rm obs})$, with $W_{\rm opt}(\mu_{\rm obs})$ determined from the Fisher matrix considerations described in Section \ref{sec:BAOsn}. Additionally, one should consider the accuracy to which $\xi_{\rm phot}(s_{\perp})$ can be modeled at high $\mu_{\rm obs}$ in order to inform any choice for a maximum $\mu_{\rm obs}$. One can then measure the angular diameter distance using $\xi_{\rm opt}(s_{\perp})$.}
\end{itemize}

We measured the clustering of mock samples intended to simulate DES Y1 data, using the approach advocated above. These measurements validated our approach and we found we were able to extract 80 per cent of the BAO information using measurements of $\xi_{\rm phot}(s_{\perp})$ and $\mu_{\rm obs}<0.8$. Our analytic and simulated results agreed that little BAO information is accessible for $\mu_{\rm obs} > 0.8$.

One important result that we obtained is that the expected precision on BAO measurements determined from $\mu_{\rm obs}$ bins of width $\Delta\mu_{\rm obs} = 0.2$ recover nearly the same uncertainty as those for $\mu_{\rm obs} < 0.8$. This is because the information between different $\mu_{\rm obs}$ is highly covariant. This suggests that measuring the BAO scale in different $\mu_{\rm obs}$ bins is likely to be an important test of the robustness of the results for the entire $\mu_{\rm obs}$ range, as disagreement might imply errors related to, e.g., the modeling assumptions or the knowledge of the photometric redshift uncertainty.

Our results show how one can construct a compressed and nearly optimal data vector to use for BAO measurements for data  with significant redshift uncertainty. We expect this to be applied to the DES Y1 data in the near future, for which we predict a 5 per cent angular diameter distance measurement to be achieved. Given we extract 80 per cent of the BAO information, we expect uncertainties can be reduced by an additional 10 per cent. As this is modest, we expect that the most significant further gains will be achieved through more precise redshift information, either via improved photometric redshifts or methods that are able to reconstruct the radial density field. Accepting the redshift uncertainty as given from a catalog, we believe this work presents a close to optimized analysis that should inform any future BAO studies conducted with purely photometric data.

\section{Acknowledgements}
We thank the anonymous referee for many helpful comments that improved the clarity of our presentation of this work.

AJR is grateful for support from the Ohio State University Center for Cosmology and ParticlePhysics.
Computer processing made use of the facilities and staff of the UK Sciama High Performance Computing cluster supported by the ICG, SEPNet and the University of Portsmouth. 
Colors made possible by 
\url{http://matplotlib.org/examples/color/named_colors.html}; figures made colorblind-friendly (hopefully) by use of Color Oracle software.

NB acknowledges the use of University of Florida's supercomputer HiPerGator 2.0 as well as thanks the University of Florida's Research Computing staff

SA thanks IFT computing facilities and support.

JGB and SA acknowledge support from the Research Project of the Spanish MINECO, Ref. FPA2015-68048-03-3P [MINECO-FEDER], and the Centro de Excelencia Severo Ochoa Program SEV-2012-0249

SA \& WJP acknowledge support from the UK Space Agency through grant ST/K00283X/1, and WJP acknowledges support from the European Research Council through grant {\it Darksurvey}, and the UK Science \& Technology Facilities Council through the consolidated grant ST/K0090X/1

Fermilab is operated by Fermi Research Alliance, LLC, under Contract No. DE-AC02-07CH11359 with the
U.S. Department of Energy. N. B. was supported by the Fermilab Graduate Student Research
Program in Theoretical Physics.

Funding for the DES Projects has been provided by the U.S. Department of Energy, the U.S. National Science Foundation, the Ministry of Science and Education of Spain, 
the Science and Technology Facilities Council of the United Kingdom, the Higher Education Funding Council for England, the National Center for Supercomputing 
Applications at the University of Illinois at Urbana-Champaign, the Kavli Institute of Cosmological Physics at the University of Chicago, 
the Center for Cosmology and Astro-Particle Physics at the Ohio State University,
the Mitchell Institute for Fundamental Physics and Astronomy at Texas A\&M University, Financiadora de Estudos e Projetos, 
Funda{\c c}{\~a}o Carlos Chagas Filho de Amparo {\`a} Pesquisa do Estado do Rio de Janeiro, Conselho Nacional de Desenvolvimento Cient{\'i}fico e Tecnol{\'o}gico and 
the Minist{\'e}rio da Ci{\^e}ncia, Tecnologia e Inova{\c c}{\~a}o, the Deutsche Forschungsgemeinschaft and the Collaborating Institutions in the Dark Energy Survey. 

The Collaborating Institutions are Argonne National Laboratory, the University of California at Santa Cruz, the University of Cambridge, Centro de Investigaciones Energ{\'e}ticas, 
Medioambientales y Tecnol{\'o}gicas-Madrid, the University of Chicago, University College London, the DES-Brazil Consortium, the University of Edinburgh, 
the Eidgen{\"o}ssische Technische Hochschule (ETH) Z{\"u}rich, 
Fermi National Accelerator Laboratory, the University of Illinois at Urbana-Champaign, the Institut de Ci{\`e}ncies de l'Espai (IEEC/CSIC), 
the Institut de F{\'i}sica d'Altes Energies, Lawrence Berkeley National Laboratory, the Ludwig-Maximilians Universit{\"a}t M{\"u}nchen and the associated Excellence Cluster Universe, 
the University of Michigan, the National Optical Astronomy Observatory, the University of Nottingham, The Ohio State University, the University of Pennsylvania, the University of Portsmouth, 
SLAC National Accelerator Laboratory, Stanford University, the University of Sussex, Texas A\&M University, and the OzDES Membership Consortium.

The DES data management system is supported by the National Science Foundation under Grant Number AST-1138766.
The DES participants from Spanish institutions are partially supported by MINECO under grants AYA2015-71825, ESP2015-88861, FPA2015-68048, SEV-2012-0234, SEV-2012-0249, and MDM-2015-0509, some of which include ERDF funds from the European Union. IFAE is partially funded by the CERCA program of the Generalitat de Catalunya.

\label{lastpage}

\end{document}